%% file: paper.tex
\documentclass[onecolumn,floatfix,superscriptaddress,showpacs,showkeys,nofootinbib]{revtex4-1}
\usepackage[T1]{fontenc}
\usepackage[latin9]{inputenc}
\usepackage{amsmath}

\usepackage[active]{srcltx} 
\usepackage{hyperref}
\usepackage{graphics}
\usepackage{epsfig}
\usepackage{amsmath}
\usepackage{mathtools}
\usepackage[english]{babel}
\usepackage{graphicx}
\usepackage{url}
\usepackage{hyperref}
 \usepackage{footnote}
\newcommand{\eq}[1]{\begin{align} #1 \end{align}}
\RequirePackage{lineno}
\usepackage{color}
\usepackage{listings}
\usepackage[normalem]{ulem}
\usepackage{ulem}

\renewcommand\sout{\bgroup\color{blue} \ULdepth=-.5ex \ULset}

\makeatletter

\newenvironment{frcseries}{\fontfamily{pzc}\selectfont}{}
\newcommand{\textfrc}[1]{{\frcseries#1}}
\newcommand{\mathfrc}[1]{\text{\textfrc{#1}}}

\makeatletter
\let\old@mathcal=\mathcal
\usepackage{dutchcal}
\let\old@dutchcal=\mathcal
\newcommand{\mathcalk}[1]{%
\if K#1
\old@mathcal{K}
\fi
\if k#1
\old@dutchcal{k}
\fi
}
\makeatother
\makeatother

\usepackage{babel}
\begin{document}
\input{macros.tex}

\global\long\def\kw{\hat{\kappa}_{2}[W]}%
\global\long\def\kk#1{\hat{\kappa}_{2}[n_{#1}]}%
\global\long\def\cc#1#2{\widehat{cov}[n_{#1},n_{#2}]}%
\global\long\def\cb#1{\bar{c}_{#1}}%
\global\long\def\kb#1{\bar{\mathfrc{k}}_{#1}}%
\global\long\def\ks#1{\mathfrc{k}_{#1}}%
\global\long\def\Cb#1{\bar{C}_{#1}}%
\global\long\def\Kb#1{\bar{\kappa}_{#1}}%

\preprint{This line only printed with preprint option}

\title{Controlling volume fluctuations for studies of critical phenomena in nuclear collisions}
\author{Romain Holzmann}
\email{r.holzmann@gsi.de}
\affiliation{GSI Helmholtzzentrum f\"ur Schwerionenforschung, 64291 Darmstadt, Germany}
\author{Volker Koch}
\email{vkoch@lbl.gov}
\affiliation{Nuclear Science Division Lawrence Berkeley National Laboratory Berkeley, CA, 94720, USA}
\affiliation{ExtreMe Matter Institute EMMI, GSI, 64291 Darmstadt, Germany}
\author{Anar Rustamov}
\email{a.rustamov@gsi.de}
\affiliation{GSI Helmholtzzentrum f\"ur Schwerionenforschung, 64291
Darmstadt, Germany}
\author{Joachim Stroth}
\email{j.stroth@gsi.de}
\affiliation{Institut f\"ur Kernphysik, Goethe-Universit\"at, 60438 Frankfurt am Main, Germany}
\affiliation{GSI Helmholtzzentrum f\"ur Schwerionenforschung, 64291 Darmstadt, Germany}
\affiliation{Helmholtz Research Academy Hesse for FAIR (HFHF), Campus Frankfurt , 60438~Frankfurt am Main, Germany}

\begin{abstract}
We generalize and extend the recently proposed method 
\cite{Rustamov:2022sqm}
to account for contributions of system size (or volume/participant) fluctuations to the experimentally measured moments of particle multiplicity distributions.  We find that in the general case there are additional biases which are not directly accessible to experiment.  These biases are, however, parametrically suppressed if the multiplicity of the particles of interest is small compared to the total charged-particle multiplicity, e.g., in the case of proton number fluctuations at top RHIC and LHC energies.  They are also small if the multiplicity distribution of charged particles per wounded nucleon is close to the Poissonian limit, which is the case at low energy nuclear collisions, e.g., at  GSI/SIS18.  We further find that mixed events are not necessarily needed to extract the correction for volume fluctuations, albeit it can help if event statistics is small, which is typically the case for reconstructing the higher-order cumulants.  We provide the formulas to correct pure and mixed cumulants of particle multiplicity distributions up to any order together with their associated biases.
\end{abstract}
\maketitle
\section{Introduction}
One of the main goals of studying relativistic heavy-ion collisions is to explore the structure of the QCD phase diagram.  
Fluctuations of observed particles carrying quantum numbers of conserved charges, baryon number (B), electric charge and strangeness, represent a powerful tool for this endeavor as the cumulants of their distributions measure the derivatives of the grand-canonical partition function, and thus the pressure ($P$), with respect to the associated chemical potentials. 
For example, for a thermal system of volume $V$ and temperature $T$, the cumulants of the net baryon number distribution, within the Grand Canonical Ensemble (GCE),
are given by~\cite{Gross:2022hyw}
\[
\kappa_{n}[B]=\frac{\partial^{n}(\ln Z)}{\partial\,(\mu_{B}/T)^{n}}=\frac{V}{T}\frac{\partial^{n}P}{\partial \,(\mu_{B}/T)^{n}},
\]
where $Z$ is a GCE partition function and $\mu_{B}$ is a baryon chemical potential.
Any nontrivial structures in the equation of state such as a possible phase transition~\cite{Stephanov:2008qz,Stephanov:2011pb,Bzdak:2019pkr,Asakawa:2009aj} will result in potentially large derivatives of the pressure and thus in large values of the cumulants of conserved charges.  
In addition, as cumulants are derivatives of the pressure, they are accessible (at vanishing or small values of chemical potential) to Lattice QCD calculations~\cite{HotQCD:2017qwq,Borsanyi:2018grb}, which in principle enables a direct comparison of results from ab initio QCD calculations with experiment. 
For example, as pointed out in Ref.~\cite{Friman:2011pf}, the measurement of higher-order cumulants close to vanishing chemical potential may test the remnants of chiral criticality. 

Measurements of fluctuations have meanwhile been carried out by many experiments. 
The STAR collaboration has measured cumulants of the net-proton number up to sixth order over the entire energy range available at RHIC~\cite{STAR:2021iop,STAR:2022etb}.
The HADES experiment has measured cumulants of proton number up to forth order at the low energy of $\sqrt{s_{\mathrm{NN}}}=2.4\gev$~\cite{Adamczewski-Musch:2020slf} and ALICE has measured the second- and third-order net-proton number cumulants at $\sqrt{s_{\mathrm{NN}}}=2.76$ and $5.02\tev$~\cite{ALICE:2019nbs, ALICE:2022xpf}.

When comparing cumulants measured in experiment with those obtained from lattice QCD or other field theoretical calculations~\cite{Fu:2016tey} one needs to be aware of several key differences.  
While theoretical calculations are typically done in the grand canonical ensemble where charges can be exchanged with a heat bath and are only conserved on the average, in experiment charges are explicitly conserved on event by event basis and one has to account for global as well as local charge conservation~\cite{Bleicher:2000ek,Bzdak:2012an,Braun-Munzinger:2018yru,Braun-Munzinger:2020jbk, Braun-Munzinger:2023gsd}. 
Also, in experiments one usually is restricted to the measurement of net protons whereas theory can only calculate cumulants of the net baryon number.  
In the presence of many pions this difference can be corrected for~\cite{Kitazawa:2011wh}. 
Finally, and this will be the topic of the present paper, in experiment the size of the particle emitting system is not constant.
Even under the tightest centrality selection criteria, this gives rise to so-called volume fluctuations \cite{Skokov:2012ds} or, equivalently, fluctuations of the number of wounded nucleons~\cite{Braun-Munzinger:2016yjz}. 
Moreover, centrality is a concept assuming a strict correlation of event activity, i.e., charged particle multiplicity in a broad region around mid rapidity, with the size of the system.
This requires a strict separation of the particles used to determine the fluctuation of conserved charges from the ones used for centrality determination. 
The effects due to volume fluctuations may be sizable, especially at lower energies where the charged particle multiplicity is dominated by the primordial protons, limiting the achievable resolution of the centrality selection.  
In Ref.~\cite{Rustamov:2022sqm} a novel and promising method based on event mixing has been proposed to experimentally determine and subtract the contributions to the cumulants caused by volume fluctuations. 
In the present work we will further elaborate on this topic, generalize the results, and provide the formulas for corrections of any higher-order cumulants.

This paper is organized as follows. 
In the next section we define the notation. 
We then present an analytical formulation of event mixing as proposed in Ref.~\cite{Rustamov:2022sqm}. 
We find that the cumulants of the the mixed events have additional bias terms which were assumed to vanish in the original work of \citep{Rustamov:2022sqm}, and we discuss the magnitude of these corrections for various scenarios. 
Next we extend our study to cumulants of higher order before
we discuss and summarize our results.
\section{Notation}
In this paper we will mostly work within the wounded-nucleon model~\cite{Bialas:1976ed}
to discuss volume or participant fluctuations.
We would like to point out that this model has its limitations when applied to low collision energies because of the moderate separation of projectile and target rapidity and a multiplicity of created particles per wounded nucleon much smaller than unity.
However, as we shall show later, the formalism can be easily applied also to the situation where one has generic volume fluctuations, as for example discussed in Refs.~\cite{Skokov:2012ds, Braun-Munzinger:2016yjz}. 
Let us start with the expression of the particle number cumulants $\kappa_j[N]$ in the presence of wounded-nucleon fluctuations (for details see Appendix~\ref{sec:Wounded-nucleon-model}): 
\begin{align}
\kappa_{1}[N] & =\ave{N_{w}}\kappa_{1}[n]=\ave{N_{w}}\ave n=\ave N\label{eq:k1}\\
\kappa_{2}[N] & =\ave{N_{w}}\kappa_{2}[n]+\ave n^{2}\kappa_{2}[N_{w}]=\Kb 2[N]+\ave N^{2}\frac{\kappa_{2}[N_{w}]}{\ave{N_{w}}^{2}}\label{eq:k2}\\
\kappa_{3}[N] & =\ave{N_{w}}\kappa_{3}[n]+3\ave n\kappa_{2}[n]\kappa_{2}[N_{w}]+\ave n^{3}\kappa_{3}[N_{w}]=\Kb 3[N]+3\ave N\Kb 2[N]\frac{\kappa_{2}[N_{w}]}{\ave{N_{w}}^{2}}+\ave N^{3}\frac{\kappa_{3}[N_{w}]}{\ave{N_{w}}^{3}}\label{eq:k3}\\
\kappa_{4}[N] & =\ave{N_{w}}\kappa_{4}[n]+4\ave n\kappa_{3}[n]\kappa_{2}[N_{w}]+3\kappa_{2}^{2}[n]\kappa_{2}[N_{w}]+6\ave n^{2}\kappa_{2}[n]\kappa_{3}[N_{w}]+\ave n^{4}\kappa_{4}[N_{w}]\nonumber \\
 & =\Kb 4[N]+4\ave N\Kb 3[N]\frac{\kappa_{2}[N_{w}]}{\ave{N_{w}}^{2}}+3\bar{\kappa}_{2}^{2}[N]\frac{\kappa_{2}[N_{w}]}{\ave{N_{w}}^{2}}+6\ave N^{2}\Kb 2[N]\frac{\kappa_{3}[N_{w}]}{\ave{N_{w}}^{3}}+\ave N^{4}\frac{\kappa_{4}[N_{w}]}{\ave{N_{w}}^{4}}\label{eq:k4}
\end{align}
Here $N$ refers to the particles of interest, say protons, and $n$ to the number of these particles arising from one wounded nucleon; thus $\ave n$ is the average number of particles per wounded nucleon.
The cumulants of the wounded-nucleon distribution are denoted by $\kappa_{j}[N_{w}]$ while the cumulants for the distribution of particles stemming from one wounded nucleon are $\kappa_{j}[n]$. The corresponding relations for cumulants of any order can be obtained with the provided software package~\cite{testGui1}.

The cumulants of interest are those at a fixed number of wounded nucleons. 
They reflect the true density fluctuations in a system at constant volume. 
We denote these cumulants for a system with fixed, i.e. non-fluctuating, number of $\ave{N_{w}}$ wounded nucleons as
\[
\Kb j[N]=\ave{N_{w}}\kappa_{j}[n],
\]

Below we will also deal with factorial cumulants, which we shall denote by $C_{j}$.
Factorial cumulants, which measure the deviation from Poisson statistics, tell us about the true correlations in the system. 
As discussed in the Appendix~\ref{sec:Cumulants-and-factorial}, they are linear combinations of the regular cumulants. 
For the first four orders we have
\begin{align*}
C_{1}[N] & =\kappa_{1}[N]=\ave N,\\
C_{2}[N] & =-\kappa_{1}[N]+\kappa_{2}[N],\\
C_{3}[N] & =2\kappa_{1}[N]-3\kappa_{2}[N]+\kappa_{3}[N],\\
C_{4}[N] & =-6\kappa_{1}[N]+11\kappa_{2}[N]-6\kappa_{3}[N]+\kappa_{4}[N].
\end{align*}
The expressions for the particle number factorial cumulants are similar to 
Eqs.~\ref{eq:k1}-~\ref{eq:k4} 
\begin{align}
C_{1}[N] & =\ave{N_{w}}C_{1}[n]=\ave{N_{w}}\ave n=\ave N,\label{eq:C1}\\
C_{2}[N] & =\Cb 2[N]+\ave N^{2}\frac{\kappa_{2}[N_{w}]}{\ave{N_{w}}^{2}},\label{eq:C2}\\
C_{3}[N] & =\Cb 3[N]+3\ave N\Cb 2[N]\frac{\kappa_{2}[N_{w}]}{\ave{N_{w}}^{2}}+\ave N^{3}\frac{\kappa_{3}[N_{w}]}{\ave{N_{w}}^{3}},\label{eq:C3}\\
C_{4}[N] & =\Cb 4[N]+4\ave N\Cb 3[N]\frac{\kappa_{2}[N_{w}]}{\ave{N_{w}}^{2}}+3\Cb 2^{2}[N]\frac{\kappa_{2}[N_{w}]}{\ave{N_{w}}^{2}}+6\ave N^{2}\Cb 2[N]\frac{\kappa_{3}[N_{w}]}{\ave{N_{w}}^{3}}+\ave N^{4}\frac{\kappa_{4}[N_{w}]}{\ave{N_{w}}^{4}}.\label{eq:C4}
\end{align}
Similar to the cumulants, we denote by 
\[
\Cb k[N]=\ave{N_{w}}C_{k}[n]
\]
the factorial cumulants for a system at constant volume or number of wounded nucleons, $\ave{N_{w}}$.

\section{Mixed events}
In Ref.~\citep{Rustamov:2022sqm} a mixed event is constructed such that it has the same total multiplicity as a given real event but each particle (track) is drawn from a different event, so that, by construction, the mixed events follow the same total multiplicity distribution as the original events. This is done in order to preserve volume flucutations as in real events.
Since each particle (track) is chosen randomly from a random event, the distribution of particle species will follow a multinomial distribution with the Bernoulli probabilities $p_{\rm i}=\ave{N_{i}}/\ave M$ for particles of type $i$. 
Here $\ave{N_{i}}$ denotes the mean number of particles of type $i$ and $\ave M$ the mean total multiplicity.
Hence, the probability to find $A$ particles (successes) of type~A and $B$ particles of type~B is given by the trinomial probability $B_{3}(A,B,M;p_{\rm A},p_{\rm B})$ and so on. 
Here $M$ denotes the multiplicity of the event under consideration. 
Thus the distribution, $P_{mix}\left(A,B\right)$, of particles of species A and B in the mixed events is obtained by folding the multiplicity distribution $P_{M}(M)$ with a trinomial (in general multinomial) distribution:
\[
P_{mix}\left(A,B\right)=\sum_{M}B_{3}(A,B,M;p_{\rm A},p_{\rm B})P_{M}(M)
\]
with 
\[
p_{\rm A}=\frac{\ave A}{\ave M},\;\;\;\;\;p_{\rm B}=\frac{\ave B}{\ave M}.
\]
and 
\begin{align}
   B_{3}(A,B,M;p_{\rm A},p_{\rm B})=\frac{M!}{A!B!(M-A-B)!}\,p_{\rm A}^{A}\,p_{\rm B}^B\,(1-p_{\rm A}-p_{\rm B})^{M-A-B}
\end{align}

The factorial-cumulant generating function for this distribution is
\begin{align}
g_{F,mx}\left(z_{\rm A},z_{\rm B}\right) & =\ln\left[\sum_{A,B}P_{mx}\left(A,B\right)(z_{\rm A})^{A}(z_{\rm B})^{B}\right]
\nonumber\\
 & =\ln\left[\sum_{M}\left[h_{3}\left(z_{\rm A},z_{\rm B}\right)\right]^{M}P_{M}(M)\right] \nonumber\\
 & =G_{F,M}\left(h_{3}\left(z_{\rm A},z_{\rm B}\right)\right)
\end{align}
where
\[
h_{3}\left(z_{\rm A},z_{\rm B}\right)=\sum_{A,B}B_{3}(A,B;M=1;p_{\rm A},p_{\rm B})(z_{\rm A})^{A}(z_{\rm B})^{B}=\left(1-p_{\rm A}-p_{\rm B}+p_{\rm A}z_{\rm A}+p_{\rm B}z_{\rm B}\right)
\]
is the factorial-moment generating function for the trinomial distribution
with one trial $\left(M=1\right)$, and $G_{F,M}\left(z\right)$ is
the factorial-cumulant generating function for the multiplicity distribution,
$P_M(M)$ (see Eq.\eqref{eq:fac_cum_gen}). The factorial cumulants are then obtained via
\[
C_{i,j}^{mix}[mix]=\left.\frac{\partial^{i}\partial^{j}}{\partial(z_{\rm A})^{i}\partial(z_{\rm B})^{j}}G_{F,M}\left(h_{3}\left(z_{\rm A},z_{\rm B}\right)\right)\right|_{z_{\rm A}=z_{\rm B}=0}=p_{\rm A}^{i}p_{\rm B}^{j}C_{i+j}\left[M\right]
\]
 with $C_{k}[M]$ being the $k^{th}$-order factorial cumulant. Using
the expression for the factorial cumulants of the multiplicity distribution
derived in Appendix \ref{sec:Multiplicity-Distribution}, Eq. \ref{eq:mult_fac_cum_wound},
we get within the wounded-nucleon model
\begin{align}
C_{1}^{mix}[A]=\kappa_{1}^{mix}[A] & =p_{\rm A}\ave{N_{w}}\ave m\nonumber \\
C_{2}^{mix}[A] & =p_{\rm A}^{2}C_{2}[M]=p_{\rm A}^{2}\left[\kappa_{2}\left[N_{w}\right]\ave m^{2}+\ave{N_{w}}C_{2}[m]\right]\nonumber \\
C_{1,1}^{mix}\left[A,B\right] & =p_{\rm A}p_{\rm B}C_{2}[M]=p_{\rm A}p_{\rm B}\left[\kappa_{2}\left[N_{w}\right]\ave m^{2}+\ave{N_{w}}C_{2}[m]\right].\label{eq:fact_cum_mix}
\end{align}
For the corresponding cumulants up to second order we get accordingly
\begin{align}
\kappa_{1}^{mix}[A]=C_{1}^{mix}[A] & =p_{\rm A}\ave{N_{w}}\ave m \nonumber\\
\kappa_{2}^{mix}[A]=C_{2}^{mix}[A]+C_{1}^{mix}[A] & =p_{\rm A}^{2}\left[\kappa_{2}\left[N_{w}\right]\ave m^{2}+\ave{N_{w}}\left(\kappa_{2}[m]-\kappa_{1}[m]\right)\right]+p_{\rm A}\ave{N_{w}}\ave m \nonumber\\
 & =p_{\rm A}^{2}\left[\kappa_{2}\left[N_{w}\right]\ave m^{2}+\ave{N_{w}}\left(\kappa_{2}[m]-\ave m\right)\right]+p_{\rm A}\ave{N_{w}}\ave m \nonumber\\
cov^{mix}\left[A,B\right]=C_{1,1}^{mix}\left[A,B\right] & =p_{\rm A}p_{\rm B}\left[\kappa_{2}\left[N_{w}\right]\ave m^{2}+\ave{N_{w}}\left(\kappa_{2}[m]-\ave m\right)\right].
\end{align}

With $\ave m$ denoting the mean number of total particles emitted by a wounded nucleon, we get $\ave a=p_{\rm A}\ave m$ and $\ave b=p_{\rm B}\ave m$ for the mean number of particles per wounded nucleon of type A and B, respectively, and recover the results of Ref.\cite{Rustamov:2022sqm}.
For that we have to assume that the multiplicity distribution per wounded nucleon is Poissonian, i.e. that $C_2[m] = \kappa_{2}[m]-\ave m=0$. 
This has been an implicit assumption in Ref.~\cite{Rustamov:2022sqm}, which however is not valid in general as we shall discuss below.

The main benefit of the event mixing is to be able to relate the factorial cumulants of the various multiplicity distributions, as can be seen from Eq.\ref{eq:fact_cum_mix}. 
All that enters is the second-order factorial cumulant, $C_{2}[M]$.
The binomial probabilities, $p_{\rm A}$ and $p_{\rm B}$, are in the sense trivial as they can be determined without any mixed events. 
Thus we may express the fluctuations of the wounded nucleons in terms of the factorial cumulant of the track multiplicity distribution 
\begin{equation}
\ave N^{2}\frac{\kappa_{2}[N_{w}]}{\ave{N_{w}}^{2}}=\frac{\ave N^{2}}{\ave M^{2}}\left(C_{2}[M]-\ave{N_{w}}C_{2}[m]\right)=\frac{\ave N^{2}}{\ave M^{2}}\left(C_{2}[M]-\Cb 2[M]\right)\label{eq:wound_fluct_2} ,
\end{equation}
where $\Cb 2[M]=\ave{N_{w}}C_{2}[m]$ is the second-order factorial cumulant for a system of $\ave{N_{w}}$ wounded nucleons \emph{without} wounded nucleon fluctuations and $N$ stands now for the multiplicity of the particles of interest, i.e., either $A$ or $B$.
While the factorial cumulant of the multiplicity distribution, $C_{2}[M]$, is accessible to experiment,
that of a non-fluctuating system, $\Cb 2[M]$, is not.  Let us, therefore
define a bias term, $\Delta_{2}$, as 
\begin{equation}
\Delta_{2}\equiv\frac{\ave N^{2}}{\ave M^{2}}\Cb 2[M].\label{eq:Delta_2}
\end{equation}
In case of a Poissonian  multiplicity distribution for one wounded nucleon the bias term vanishes, i.e., $\Delta_{2}=0$, since $\Cb 2[M]=\ave{N_{w}}C_{2}[m]=0$ in this case, and we recover the results of Ref.~\cite{Rustamov:2022sqm}.
Let us furthermore define the corrected cumulant, $\kappa_{2}^{corr}[N],$ which is based  on measurable quantities only
\begin{equation}
\kappa_{2}^{corr}[N]=\kappa_{2}[N]-\frac{\ave N^{2}}{\ave M^{2}}C_{2}[M].\label{eq:k2_corr_1}
\end{equation}
Following Eq. \ref{eq:k2} and using Eq.~\ref{eq:C2}, the cumulant of  the system \emph{without} wounded nucleon fluctuations, $\Kb 2[N]$, is given by
\begin{align}
\Kb 2[N]=\kappa_{2}[N]-\ave N^{2}\frac{\kappa_{2}[N_{W}]}{\ave{N_{W}}^{2}} & =\kappa_{2}^{corr}[N]+\Delta_{2}.
\label{eq:k2_bar_1}
\end{align}

The bias, $\Delta_{2}$, while not directly measurable, may be constrained by a fit to the track multiplicity distribution within the wounded-nucleon model~\cite{Bialas:1976ed}, as it is commonly done~\cite{HADES:2017def, STAR:2022etb, ALICE:2013hur}.
In addition, we note that for protons at very high collision energies we have $\ave{N_{p}}\ll\ave M$ so that $\Delta_{2}$ is suppressed parametrically.
This behavior can indeed be illustrated with simulations as presented in Sec.~\ref{sec:Simulations}.
Since cumulants scale with the system size, or in our case with the number of wounded nucleons, $\ave{N_{w}}$, it is instructive to scale the (factorial) cumulants with the mean number of particles
\begin{equation}
\frac{\Kb 2[N]}{\ave N} =\frac{\kappa_2[N]}{\ave N} -\frac{\ave N}{\ave M}\left(\frac{C_{2}[M]}{\ave N} -\frac{\Cb 2[M]}{\ave N} \right)=\frac{\kappa_2^{corr}[N]}{\ave N}+\frac{\Delta_{2}}{\ave N}
\end{equation}
The scaled bias is then given by 
\begin{equation}
\frac{\Delta_{2}}{\ave N}=\frac{\ave N}{\ave M}\cb 2[M].\label{eq:delta_2}
\end{equation}
Typically, the scaled cumulants are of order unity, $\kappa_j[N]/\ave{N}\sim\mathcal{O}(1)$. 
In addition, the scaled factorial cumulants, $C_k[N]/\ave N$, are expected to depend only weakly on the multiplicity. 
Therefore, the scaled bias should be much smaller than one, $\Delta_{2}/\ave{N}\ll1$, for the volume correction to be reliable.

Finally, one may express the bias term $\Delta_{2}$ 
also in terms of cumulants by using the relation between cumulants
and factorial cumulants (see Appendix \ref{sec:Cumulants-and-factorial}), $C_{2}[M]=\kappa_{2}[M]-\ave M$ and and so forth. 
This gives,
\begin{align}
\Delta_{2} & =\frac{\ave N^{2}}{\ave M^{2}}\left(\Kb 2[M]-\ave M\right)\label{eq:Delta_2_kappa}
\end{align}
A note of caution may be useful in this context. 
One might be inclined to express the fluctuations of the wounded nucleon directly using the cumulants of the multiplicity distribution, in which case one would get
\[
\ave N^{2}\frac{\kappa_{2}[N_{w}]}{\ave{N_{w}}^{2}}=\frac{\ave N^{2}}{\ave M^{2}}\left(\kappa_{2}[M]-\ave{N_{w}}\kappa_{2}[m]\right).
\]
And since $\ave{N_{W}}\kappa_{2}[m]=\Kb 2[M]$ is not directly accessible
to experiment, one may further assign the bias to be $\Delta_{2}=\frac{\ave N^{2}}{\ave M^{2}}\Kb 2[M]$.
This, however, would considerably overestimate its true value, Eq. \ref{eq:Delta_2_kappa}, as cumulants always contain a ``trivial'' component proportional to the number of particles, which in principle is measurable. 
\section{Higher-Order Results}\label{sec:HigherOrders}
Let us now discuss the corrections for volume fluctuations up to fourth order. 
Given the discussion in the previous section the strategy is straightforward. 
First we express the fluctuations of the wounded nucleons in terms of factorial cumulants of the multiplicity distribution. 
Then we identify the parts which are experimentally accessible and those which are not. 
The latter will be the bias while the former will be subtracted from the expression for the cumulants in order to remove most of the effect of volume fluctuations. 
The terms involving cumulants of the wounded-nucleon distribution as they appear in the expressions for the cumulants as $\kappa_{j}[N_{w}]/\ave{N_{w}}^{j}$, see Eqs. (\ref{eq:C2}-\ref{eq:C4}) are:
\begin{align}
\frac{\kappa_{2}[N_{w}]}{\ave{N_{w}}^{2}} & =\frac{C_{2}[M]-\bar{C}_{2}[M]}{\langle M\rangle^{2}}\label{eq:k_wound_2}\\
\frac{\kappa_{3}[N_{w}]}{\ave{N_{w}}^{3}} & =-3\frac{\bar{C}_{2}[M]}{\langle M\rangle^{2}}\frac{\kappa_{2}[N_{w}]}{\ave{N_{w}}^{2}}+\frac{C_{3}[M]-\bar{C}_{3}[M]}{\langle M\rangle^{3}}\label{eq:k_wound_3}\\
\frac{\kappa_{4}[N_{w}]}{\ave{N_{w}}^{4}} & =-6\frac{\bar{C}_{2}[M]}{\langle M\rangle^{2}}\frac{\kappa_{3}[N_{w}]}{\ave{N_{w}}^{3}}-\frac{4\bar{C}_{3}[M]\langle M\rangle+3\bar{C}_{2}[M]{}^{2}}{\langle M\rangle^{4}}\frac{\kappa_{2}[N_{w}]}{\ave{N_{w}}^{2}}+\frac{C_{4}[M]-\bar{C}_{4}[M]}{\langle M\rangle^{4}} \label{eq:k_wound_4}
\end{align}
We note that binomial efficiency corrections do not affect the results as both, numerators and denominators of the right hand side of the above expressions, scale with the same power of the efficiency.

Inserting these expressions into Eqs.~(\ref{eq:k2}-\ref{eq:k4}) for the cumulants $\kappa_j[N]$, we can solve for the cumulants of the system with fixed number of wounded nucleons, namely the $\Kb j[N]$. 
The results are given in the following general form

\begin{equation}
    \Kb j[N]=\kappa_{j}^{corr}[N]+\Delta_{j}[N]
    \label{eq_kappa_bar}
\end{equation}
with $\Kb j[N]$ the cumulant of order $j$ for a system with fixed $N_{w}$ nucleons, $\kappa_{j}^{corr}[N]$ the cumulant including the measurable corrections for volume fluctuations, and $\Delta_{j}$ the corresponding bias due to quantities that are not measurable.
The second-order result we already derived in Sec.~\ref{sec:Wounded-nucleon-model},
Eqs. (\ref{eq:k2_corr_1}) and (\ref{eq:Delta_2}), namely 
\begin{align}
\kappa_{2}^{corr}[N] & =\kappa_{2}[N]-\frac{\ave N^{2}}{\ave M^{2}}C_2[M]\nonumber \\
\Delta_{2} & =\frac{\ave N^{2}}{\ave M^{2}}\Cb 2[M].\label{eq:k_corr_2}
\end{align}
For the third order we have
\begin{align}
\kappa_{3}^{corr}[N] & =\kappa_{3}[N]-\frac{3C_{2}[M]\kappa_{2}[N]\langle N\rangle}{\langle M\rangle^{2}}+\frac{3C_{2}[M]{}^{2}\langle N\rangle^{3}}{\langle M\rangle^{4}}-\frac{C_{3}[M]\langle N\rangle^{3}}{\langle M\rangle^{3}}\nonumber \\
\Delta_{3} & =\bar{C}_{2}[M]\left(\frac{3\kappa_{2}[N]\langle N\rangle}{\langle M\rangle^{2}}-\frac{3C_{2}[M]\langle N\rangle^{3}}{\langle M\rangle^{4}}\right)+\frac{\bar{C}_{3}[M]\langle N\rangle^{3}}{\langle M\rangle^{3}}.\label{eq:k_corr_3}
\end{align}
And the fourth order result reads
\begin{align}
\kappa_{4}^{corr}[N] & =\kappa_{4}[N]-\left(\frac{6\kappa_{2}[N]\langle N\rangle^{2}\left(C_{3}[M]\langle M\rangle-3C_{2}[M]{}^{2}\right)}{\langle M\rangle^{4}}+\frac{4C_{2}[M]\kappa_{3}[N]\langle N\rangle}{\langle M\rangle^{2}}\right.\nonumber \\
 & \left.+\frac{3C_{2}[M]\kappa_{2}(N){}^{2}}{\langle M\rangle^{2}}+\frac{\langle N\rangle^{4}\left(-10C_{3}[M]C_{2}[M]\langle M\rangle+15C_{2}[M]{}^{3}\right)}{\langle M\rangle^{6}}+\frac{C_{4}[M]\langle N\rangle^{4}}{\ave M^{4}}\right)\nonumber \\
\Delta_{4} & =\bar{C}_{2}[M]\left(-\frac{18C_{2}[M]\kappa_{2}[N]\langle N\rangle^{2}}{\langle M\rangle^{4}}+\frac{15C_{2}[M]{}^{2}\langle N\rangle^{4}}{\langle M\rangle^{6}}-\frac{4C_{3}[M]\langle N\rangle^{4}}{\langle M\rangle^{5}}+\frac{4\kappa_{3}[N]\langle N\rangle}{\langle M\rangle^{2}}+\frac{3\kappa_{2}[N]{}^{2}}{\langle M\rangle^{2}}\right)\nonumber \\
 & +\bar{C}_{3}[M]\left(\frac{6\kappa_{2}[N]\langle N\rangle^{2}}{\langle M\rangle^{3}}-\frac{6C_{2}[M]\langle N\rangle^{4}}{\langle M\rangle^{5}}\right)+\frac{\bar{C}_{4}[M]\langle N\rangle^{4}}{\langle M\rangle^{4}}.\label{eq:k_corr_4}
\end{align}

The corresponding relations for correction and bias terms of any order can be obtained with the provided software package~\cite{testGui1}.

Equivalent expressions for the factorial cumulants, $C_{n}[N]$, and their related biases, $\Delta_{n,F}$, may then be obtained by using the relation between factorial cumulants and regular cumulants Eq.~\ref{eq:fac_cum_fom_cum_Bell}:
\begin{align*}
C_{n}^{corr}[N] & =\sum_{j=1}^{n}B_{n,j}\left(1,-1,2,\ldots,(-1)^{j-1}(n-j+1)!\right)\,\kappa^{corr}_{j}[N]\\
\Delta_{n,F} & =\sum_{j=1}^{n}B_{n,j}\left(1,-1,2,\ldots,(-1)^{j-1}(n-j+1)!\right)\,\Delta_{j}[N]
\end{align*}
The results for the corrected factorial cumulants, $C_k^{corr}$ and $c_k^{corr}$, and the associated biases, $\Delta_{k,F}$ and $\delta_{k,F}$, are given in  Appendix \ref{sec:results_factorial_cumulants}.

\section{Simulations}\label{sec:Simulations}
Experimental data are usually analyzed in centrality percentiles, i.e.\ event classes corresponding to the $n\%$ most central collisions, by introducing selection criteria on e.g.\ the energy deposited in a forward detector system covering typically the projectile (target) spectator region or the multiplicity of charged particles emitted from the mid-rapidity region, with an acceptance reaching close to the projectile (target) rapidity regions in case of low beam energies~\cite{ALICE:2013hur, HADES:2017def}. 
For the latter, care must be taken to ensure that the evaluated particles are not simultaneously used to determine the critical fluctuations~\cite{STAR:2020tga}.
\begin{figure}[!htb]
    \centering
    \includegraphics[width=.4\linewidth,clip=true]{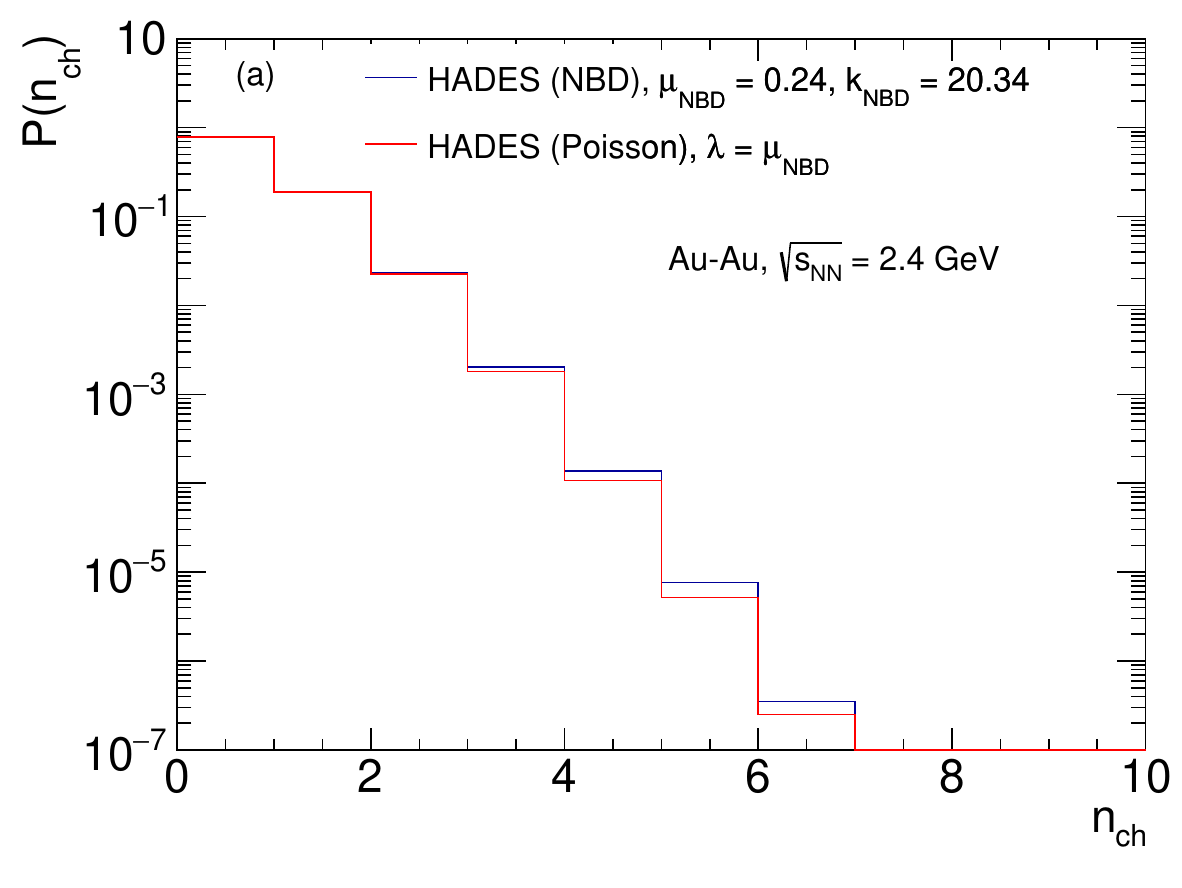}
     \includegraphics[width=.4\linewidth,clip=true]{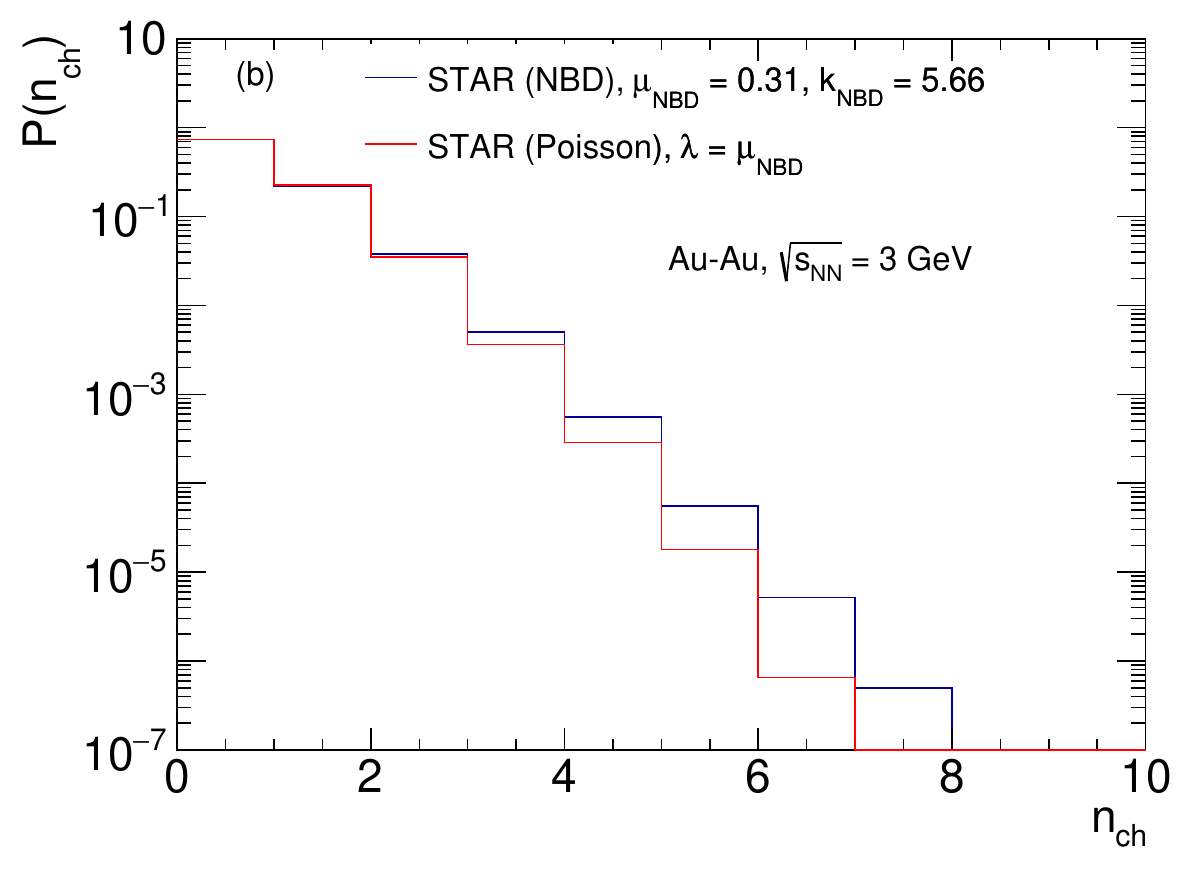}
     \includegraphics[width=.4\linewidth,clip=true]{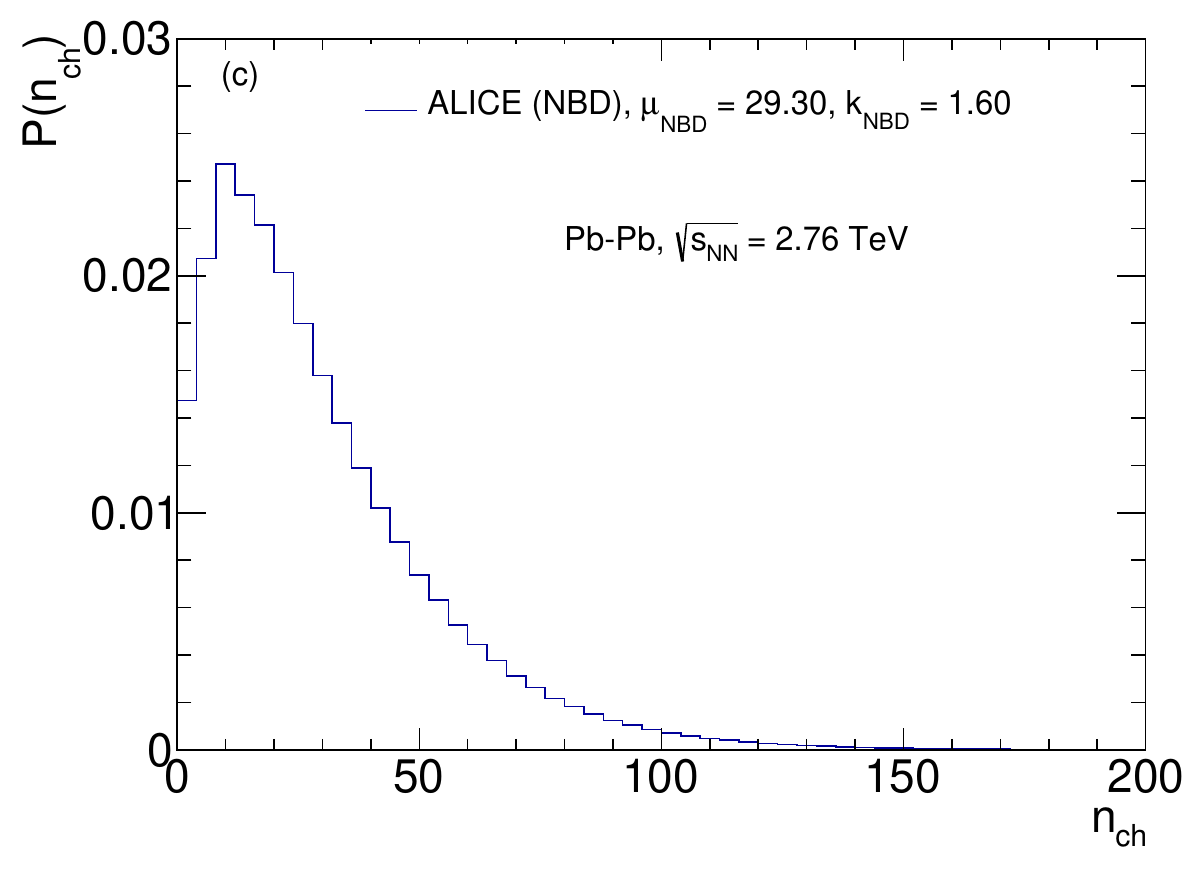}
    \caption{NBD adjusted to HADES (a), STAR (b) and ALICE (c) charged-particle multiplicity distributions (blue line), shown together with a Poisson distribution for the HADES and STAR data (red lines), the  parameters used are listed in Table I.}
    \label{fig:NBD_HADES_STAR}
\end{figure}
\begin{figure}[!htb]
    \centering
    \includegraphics[width=.4\linewidth,clip=true]{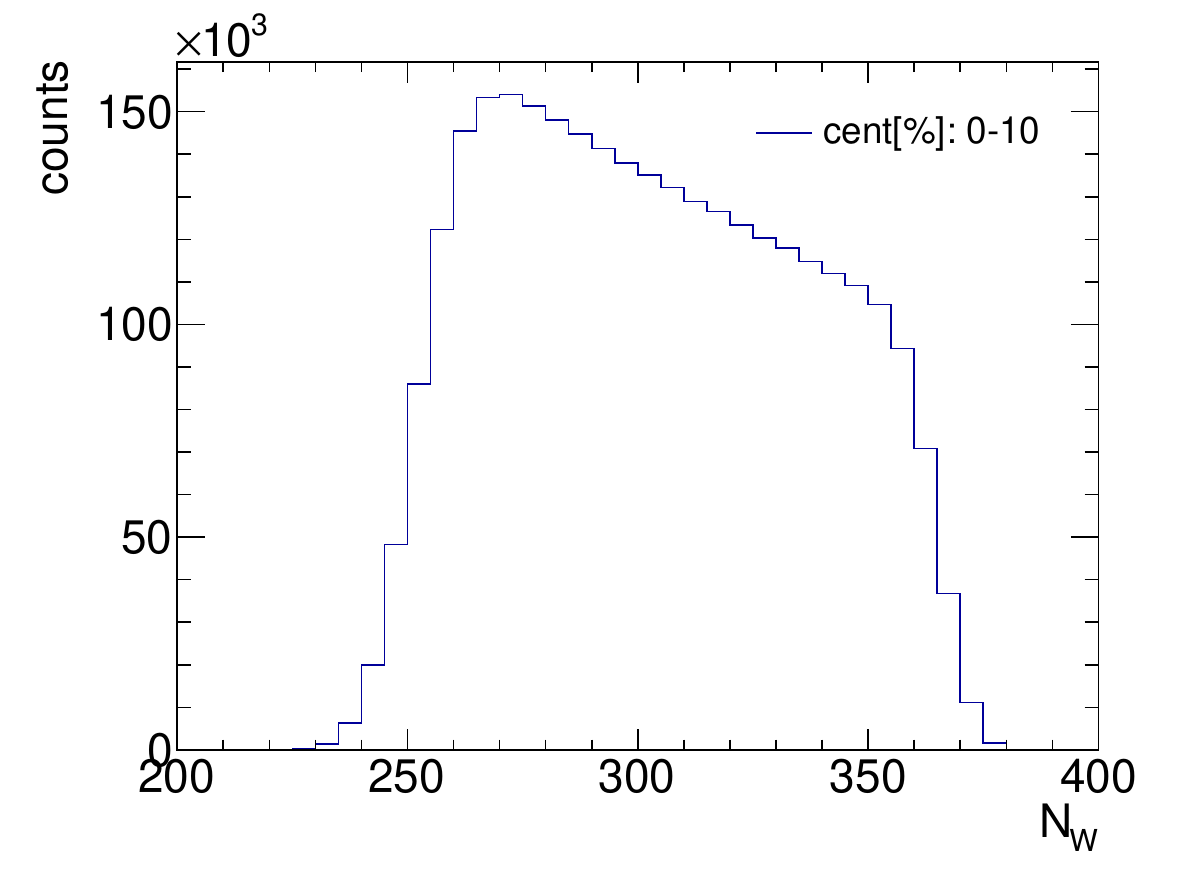}
     \includegraphics[width=.4\linewidth,clip=true]{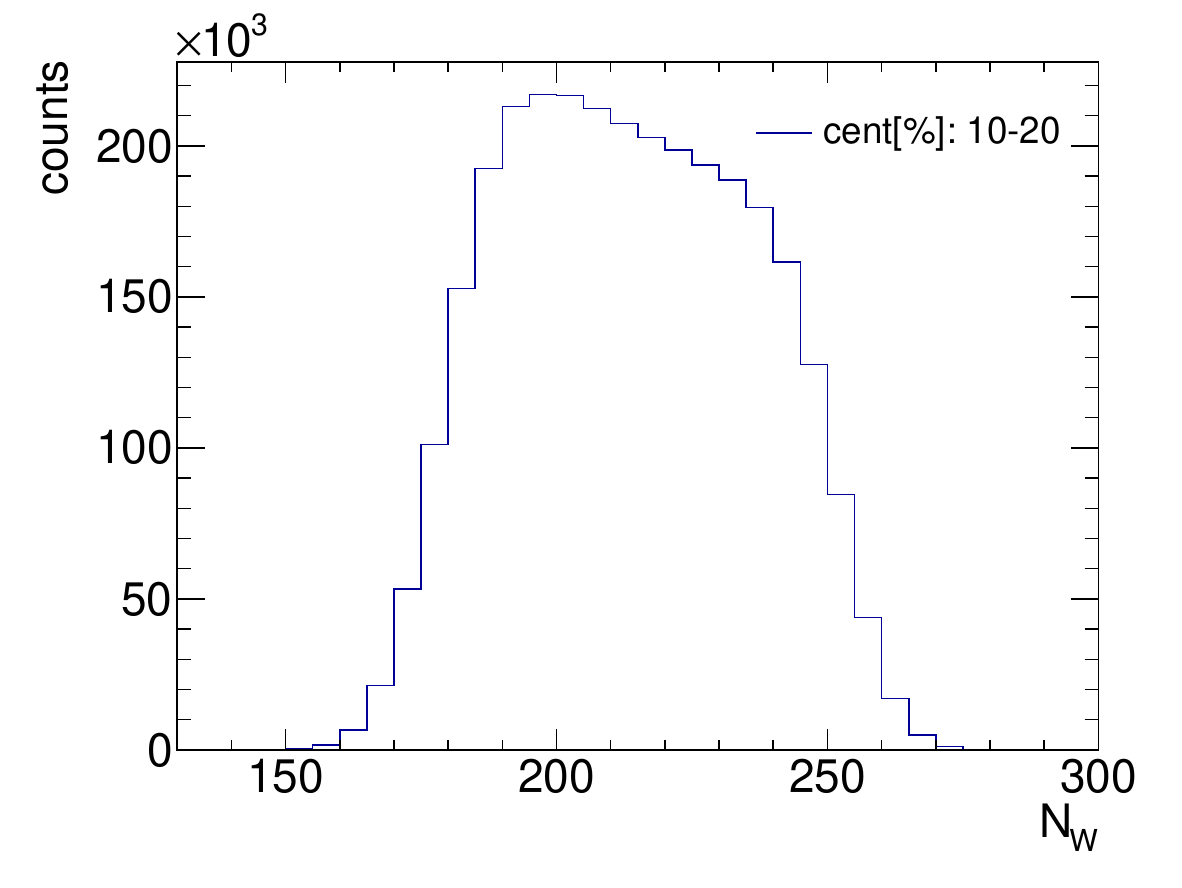}
     \includegraphics[width=.4\linewidth,clip=true]{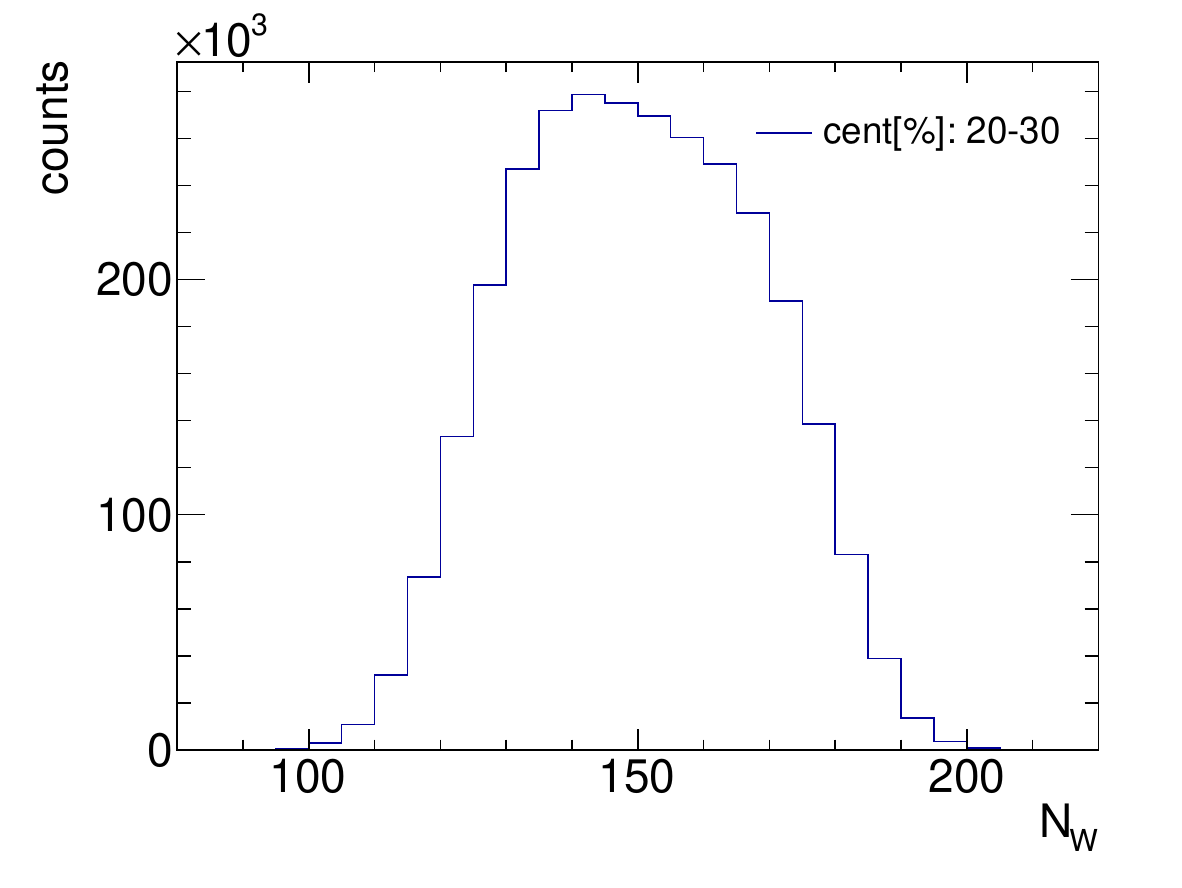}
     \includegraphics[width=.4\linewidth,clip=true]{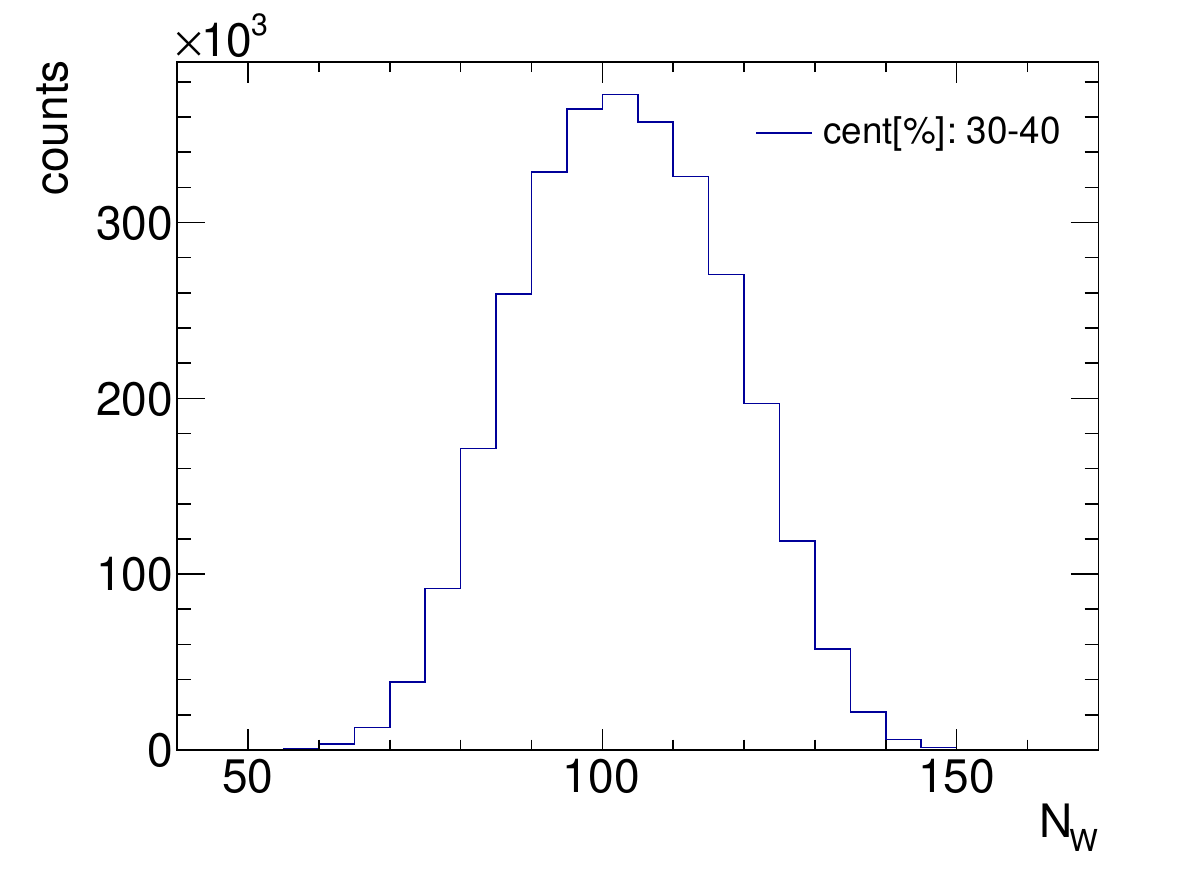}
    \caption{Distribution of wounded nucleons in Au+Au collisions at $\sqrt{s_{\mathrm{NN}}}$ = 2.4 GeV for four selected centrality classes, as obtained from the  Glauber Monte Carlo simulations.}
    \label{fig:NW_dist}
\end{figure}
The respective distributions, like e.g., the forward energy deposit or the charged particle multiplicity, are commonly modelled using the Glauber Monte Carlo Model~\cite{Loizides:2014vua}. 
The model provides event by event and for a given impact parameter the number of projectile/target nucleons which are ``wounded'' and responsible for the event activity (multiplicity), and those, which proceed nearly undisturbed into the phase space region covered by the forward detectors. 
To determine centrality using  charged particle multiplicity the respective distribution is generally modelled assuming that particles are ``produced'' independently from  distinct sources following a negative binomial distribution (NBD). 
Its probability mass function is defined as
\begin{equation}
P(n;\mu,k)=\frac{\Gamma(n+k)}{\Gamma(n+1)\Gamma(k)}\left(\frac{\mu}{k}\right)^{n}\left(\frac{\mu}{k}+1 \right)^{-(n+k)},
\end{equation}
where $\mu$ denotes the mean of the NBD, while the combination of $\mu$ and $k$ determines its higher-order cumulants
\begin{equation}
\kappa_{n}^{NBD}=\left. \frac{\partial^{n} \ln M(t)}{\partial t^{n}}\right\rvert_{t = 0},
\label{nbd_cums}
\end{equation}
with 
\begin{equation}
M(t)= \sum_{n=0}^{\infty} e^{tn}P(n;\mu,k)=\left(\frac{k}{k + \mu(1 - e^{t})}\right)^{k}
\end{equation}
%
being the moment-generating function of the NBD.  
The first four cumulants read
\eq{
\label{cum1}
&\kappa_{1}^\text{NBD} = \mu,\\
\label{cum2}
&\kappa_{2}^\text{NBD}= \frac{\mu(k+\mu)}{k},\\
\label{cum3}
&\kappa_{3}^\text{NBD} = \frac{\mu(k+\mu)(k+2\mu)}{k^{2}},\\
\label{cum4}
&\kappa_{4}^\text{NBD} = \frac{\mu(k+\mu)(k^{2}+6k\mu+6\mu^{2})}{k^3},
}
The parameters of the NBD are fixed in each experiment by the fitting procedure.
In a first step the number of particle-emitting sources $n_{s}$ is determined according to~\cite{ALICE:2013hur, STAR:2022etb}
\begin{equation}
n_\text{s} = fN_{w} + (1-f)N_{coll},
\label{eqn_sources}
\end{equation}
where $N_{W}$ and $N_{coll}$ are the numbers of wounded nucleons and binary collisions, respectively. 
Sampling impact parameters according to $d\sigma = b\,db$ a list of number of sourses,$n_\text{s}^i$, is generated, with $i\in [1, \cdots, N_\text{Event}]$.
Then, for each event $i$, NBD is sampled $n_\text{s}^i$ times and the parameters of the NBD, $\mu$, $k$ and  $f$, are adjusted such that the obtained multiplicity distribution agrees with the corresponding experimental one.
The mixing parameter $f$ is introduced to improve the description by accounting also particles produced in hard (prompt) processes.
\begin{figure}[!htb]
    \centering
    \includegraphics[width=.6\linewidth,clip=true]{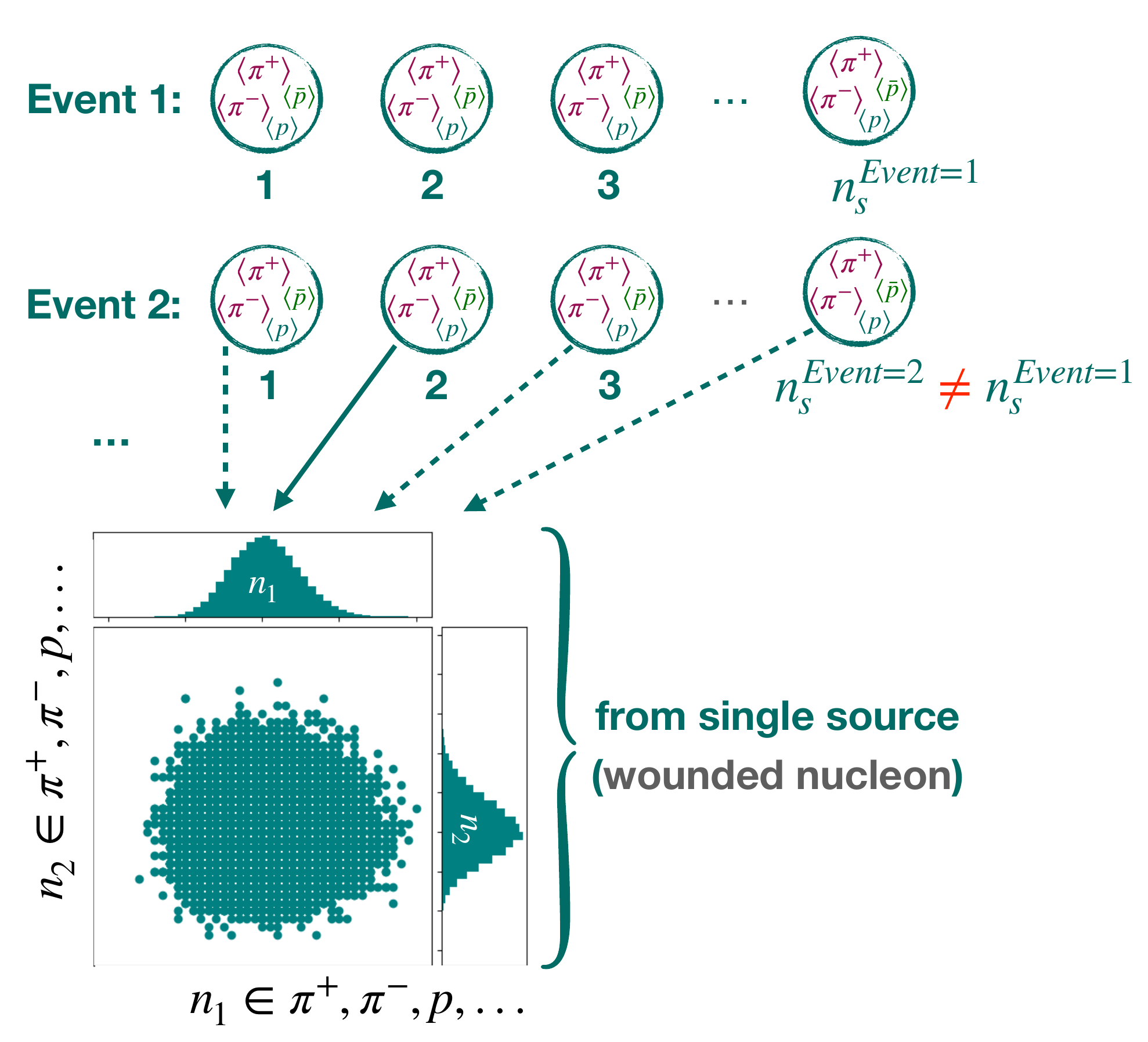}
    \caption{Simulation of particle production within the model of independent sources. The circles indicate the individual sources sampled according to Eq.~\ref{eqn_sources}. $n_1$ and $n_2$ show particle species used to sample emission from a single source. The distributions per single source can be chosen arbitrarily. See models A and B discussed below.}
    \label{fig:simulations}
\end{figure}

\begin{figure}[!htb]
    \centering
    \includegraphics[width=.45\linewidth,clip=true]{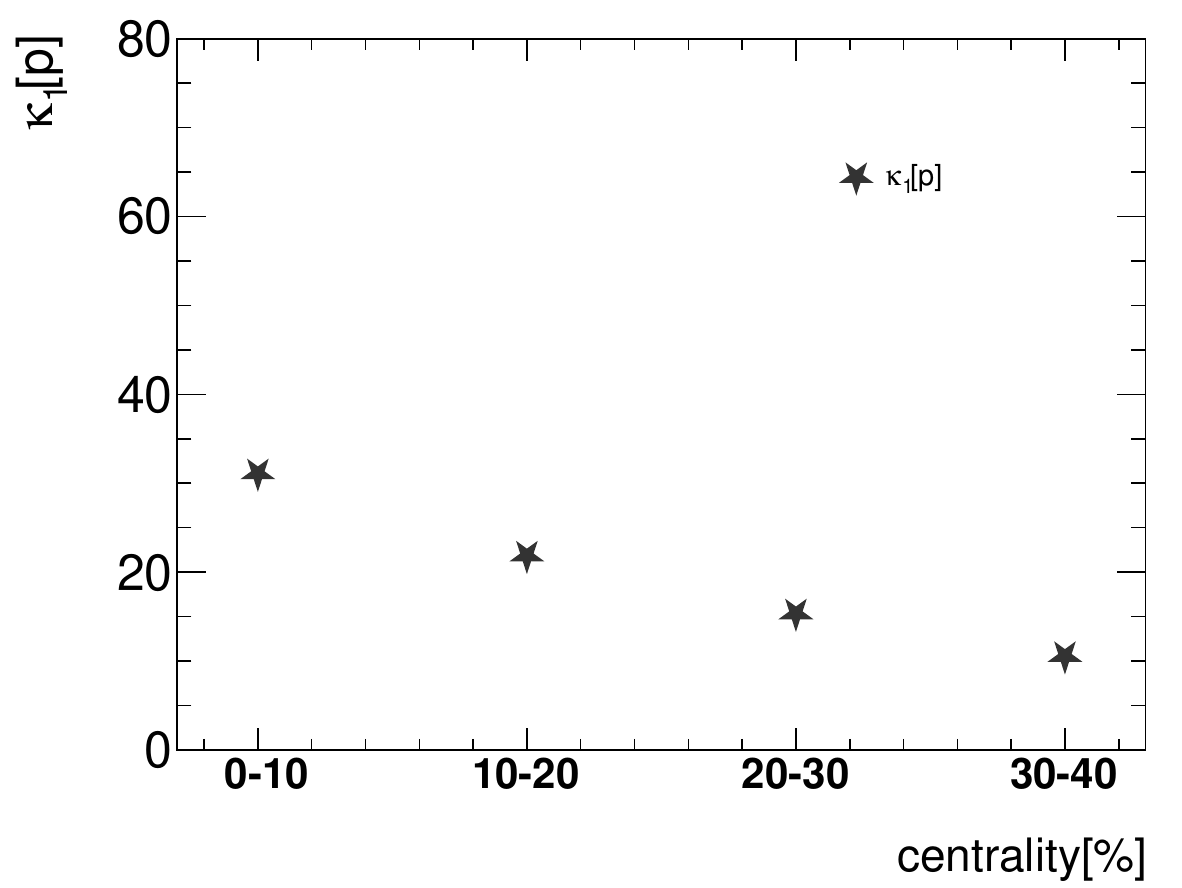}
     \includegraphics[width=.45\linewidth,clip=true]{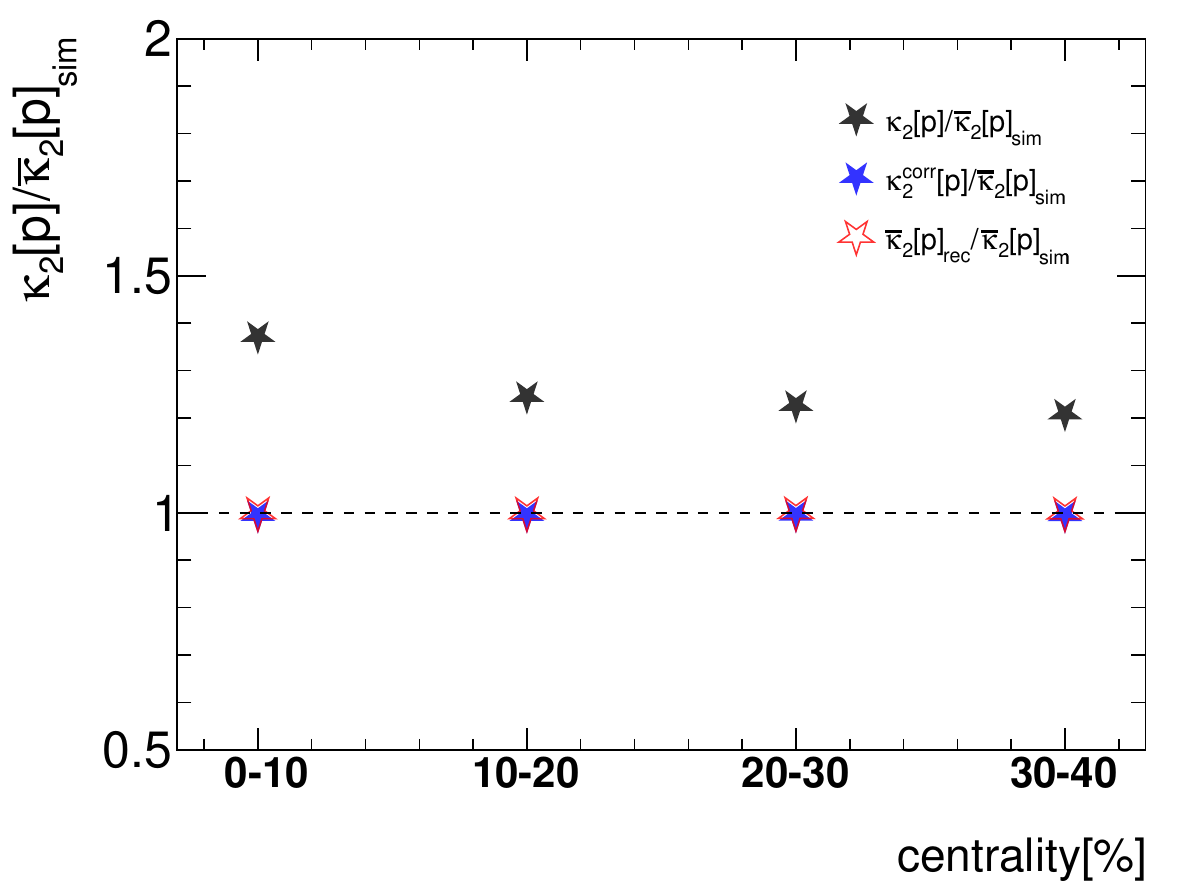}
      \caption{Left panel: Mean number of simulated protons used in Model A as a function of centrality. Right panel: Second-order cumulants of protons in Model A including volume fluctuations (black stars), corrected with Eq.~\ref{eq:k_corr_2} (blue stars) and reconstructed with Eq.~\ref{eq_kappa_bar} (open red stars). The results are normalized to $\bar{\kappa}_{2}[p]_{sim}$, corresponding to the second-order cumulants of protons in the absence of volume fluctuations.}
    \label{fig:kappa1_2Protons}
\end{figure}
\begin{figure}[!htb]
    \centering
    \includegraphics[width=.45\linewidth,clip=true]{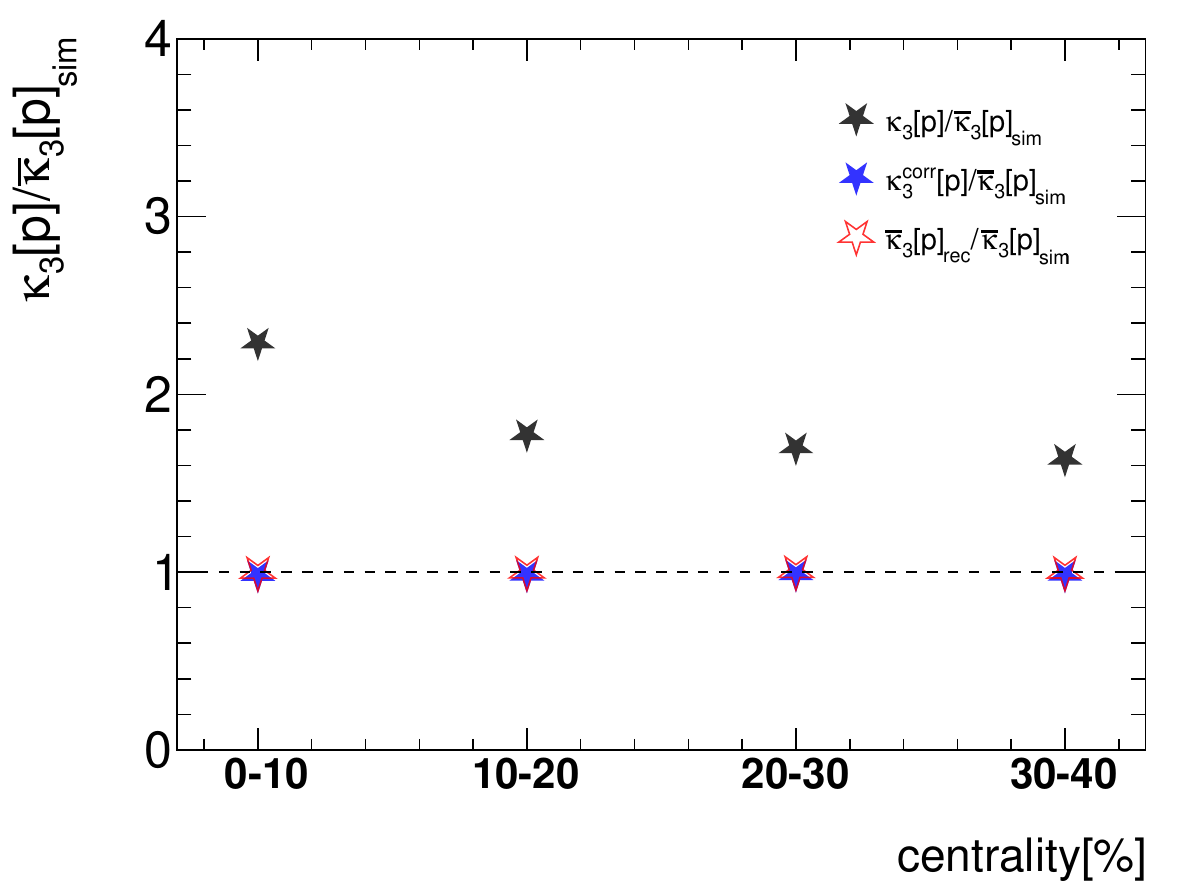}
     \includegraphics[width=.45\linewidth,clip=true]{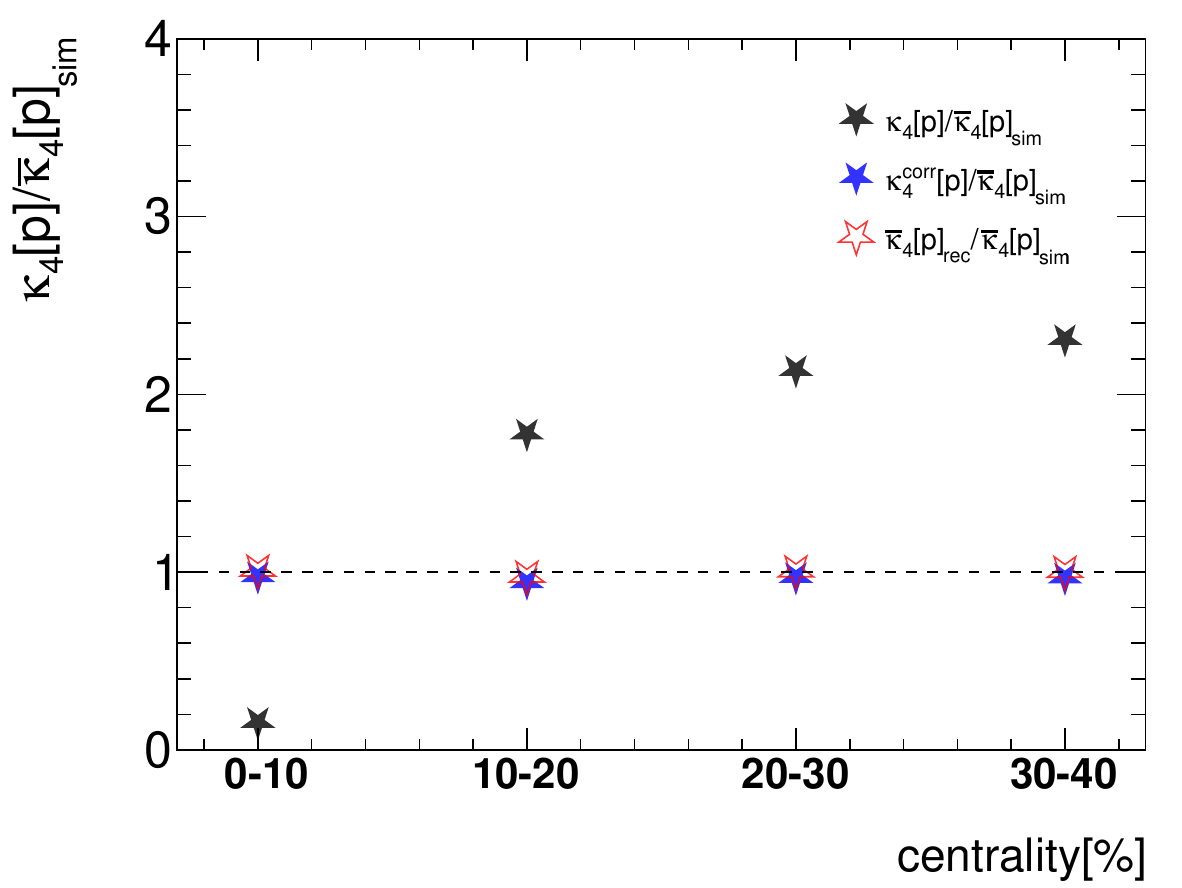}
      \caption{Left panel:  Third order cumulants of protons in Model A including volume fluctuations (black stars), corrected with Eq.~\ref{eq:k_corr_3} (blue stars) and reconstructed with Eq.~\ref{eq_kappa_bar} (open red stars). The results are normalized to $\bar{\kappa}_{3}[p]_\text{sim}$, corresponding to the second-order cumulants of protons in the absence of volume fluctuations. Right panel:  Fourth order cumulants of protons in Model A including volume fluctuations (black stars), corrected with Eq.~\ref{eq:k_corr_4} (blue stars) and reconstructed with Eq.~\ref{eq_kappa_bar} (open red stars). The results are normalized to $\bar{\kappa}_{4}[p]_\text{sim}$, corresponding to the fourth order cumulants of protons in the absence of volume fluctuations. }
    \label{fig:kappa3_4Protons}
\end{figure}
\begin{figure}[!htb]
    \centering
    \includegraphics[width=.45\linewidth,clip=true]{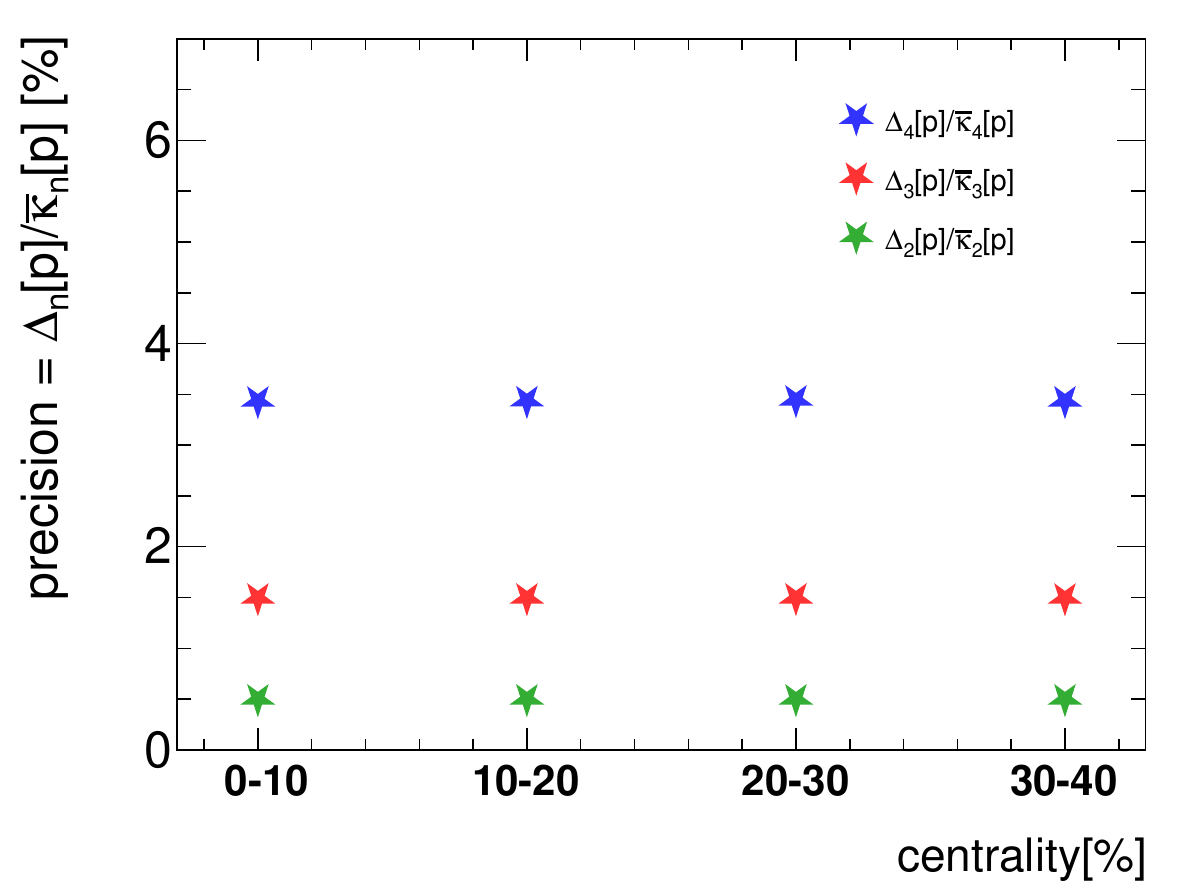}
      \caption{The normalised bias terms in Model A.}
    \label{fig:bias_model_A}
\end{figure}
Panel (a) of Fig.~\ref{fig:NBD_HADES_STAR} represents the NBD distribution as observed by the HADES experiment for Au+Au collisions at $\sqrt{s_\mathrm{NN}} = 2.4$\,GeV, with parameters $\mu = 0.24$, $k = 20.34$, and $f = 1$ taken from~\cite{HADES:2017def}. 
For comparison, we also present a Poisson distribution with the same mean, $\mu$.  
Similar distributions from the STAR~\cite{STAR:2022etb} and ALICE~\cite{ALICE:2013hur} experiments are presented in panels (b) and (c).
Figure~\ref{fig:NBD_HADES_STAR} shows that at the HADES energy the fitted NBDs  are very close to Poisson distributions. 
Quantitatively this can be seen by evaluating the cumulants of the HADES NBD ($\mu = 0.24$, $k = 20.34$)
\eq{
\label{cum1n}
&\kappa_{1}^\text{NBD}(\text{HADES}) = 0.24 \\
\label{cum2n}
&\kappa_{2}^\text{NBD}(\text{HADES}) = 0.2428,\\
\label{cum3n}
&\kappa_{3}^\text{NBD}(\text{HADES}) = 0.2486,\\
\label{cum4n}
&\kappa_{4}^\text{NBD}(\text{HADES})=  0.2602.
}
For a Poisson distribution all cumulants are equal to its mean and the HADES data are indeed close to fulfilling this condition. 
The statement, to a lesser extent, is also valid for the STAR Au+Au data at 3\,GeV (see Fig.~\ref{fig:NBD_HADES_STAR}). 
The corresponding ALICE distribution, however, is much wider compared to the Poisson baseline, but the ALICE NBD is obtained for very different acceptance than that used for the cumulant analysis.
In Table~\ref{tab:NBDparameters} we also provide the NBD parameters as obtained by the STAR and ALICE collaborations for Au+Au and Pb+Pb collisions at $\sqrt{s_{\mathrm{NN}}} = 3$\,GeV and 2.76\,TeV, respectively.
 
\begin{table}[!htbp]
\centering
 \begin{tabular}{||c | c | c | c||} 
 \hline
experiments & $\mu$ & $k$ & $f$  \\ [0.5ex] 
 \hline
 HADES & 0.24 & 20.34 & 1  \\ 
 \hline
 STAR & 0.31 & 5.66 & 0.94 \\
 \hline
 ALICE & 29.3 & 1.6 & 0.8\\
\hline
\end{tabular}
\caption{\label{tab:NBDparameters} NBD parameters as extracted from Glauber fits to particle distributions observed in different experiments.  
For simulations, the distributions should be folded within the experimental acceptance in which the cumulants are measured.}
\end{table}

In the following we test the proposed method using two different simulations referred to as Model~A and Model~B. 
While the sampling of wounded nucleons is the same for both models, in Model~A we sample different particle species independently while in Model~B we introduce correlations between pions and protons via cluster production and decay. 
We will concentrate on the HADES data. 
For both methods, we use the Glauber model to extract the distributions of wounded nucleons corresponding to four different centrality classes in Au+Au collisions at $\sqrt{s_{\mathrm{NN}}} = 2.4$\,GeV.  
They are presented in Figure~\ref{fig:NW_dist}.

In both models particles are produced from independent sources (cf.\ Eq.~\ref{eqn_sources}). 
For the HADES data the extracted number of binary collision is zero ($f=1$, see Table~\ref{tab:NBDparameters} ), the number of sources per event are sampled exclusively from the wounded nucleon distributions presented in Fig.~\ref{fig:NW_dist}. 
The simulation process is schematically illustrated in Fig.~\ref{fig:simulations}. 

\begin{figure}[!htb]
    \centering
    \includegraphics[width=.45\linewidth,clip=true]{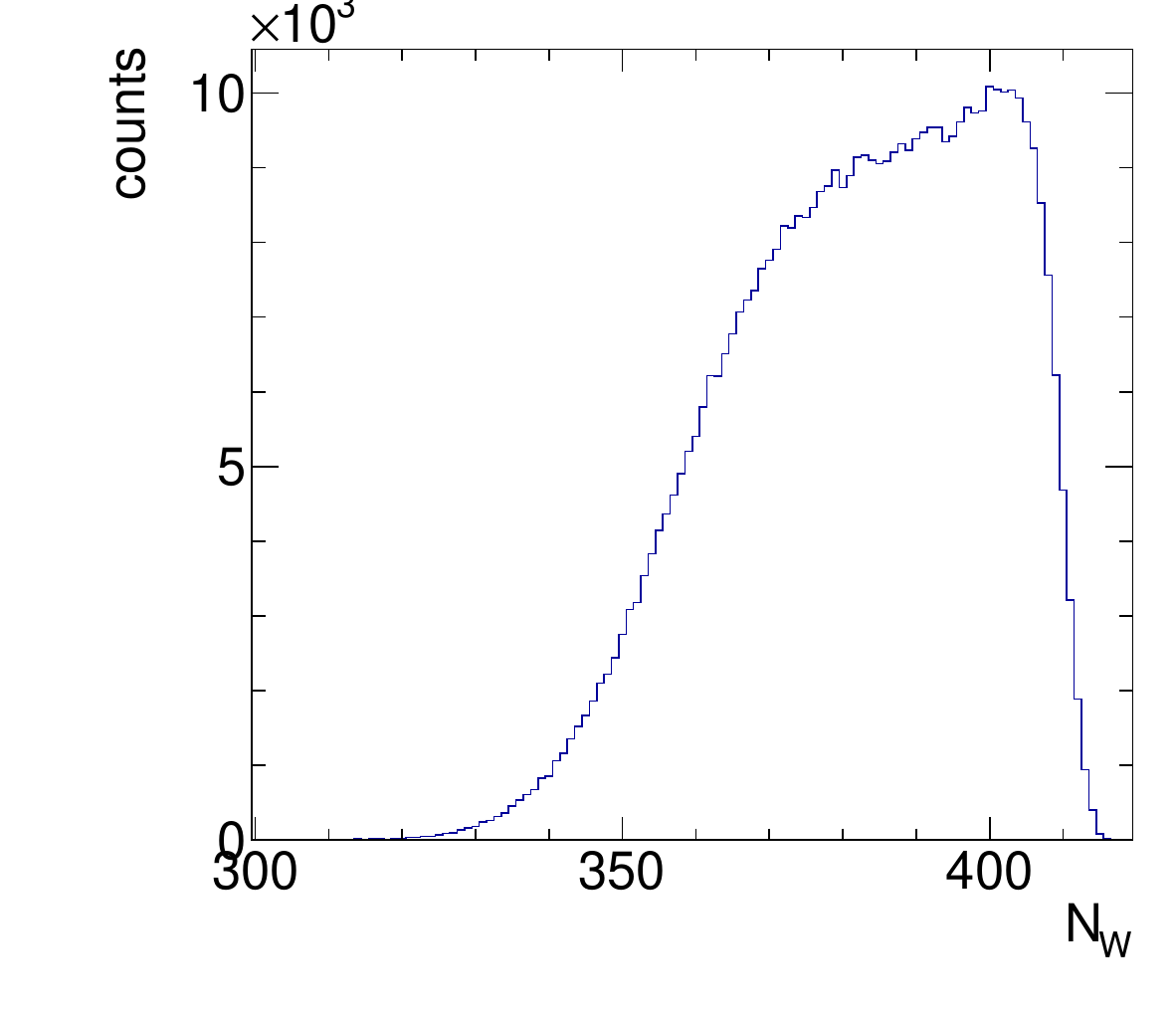}
    \includegraphics[width=.45\linewidth,clip=true]{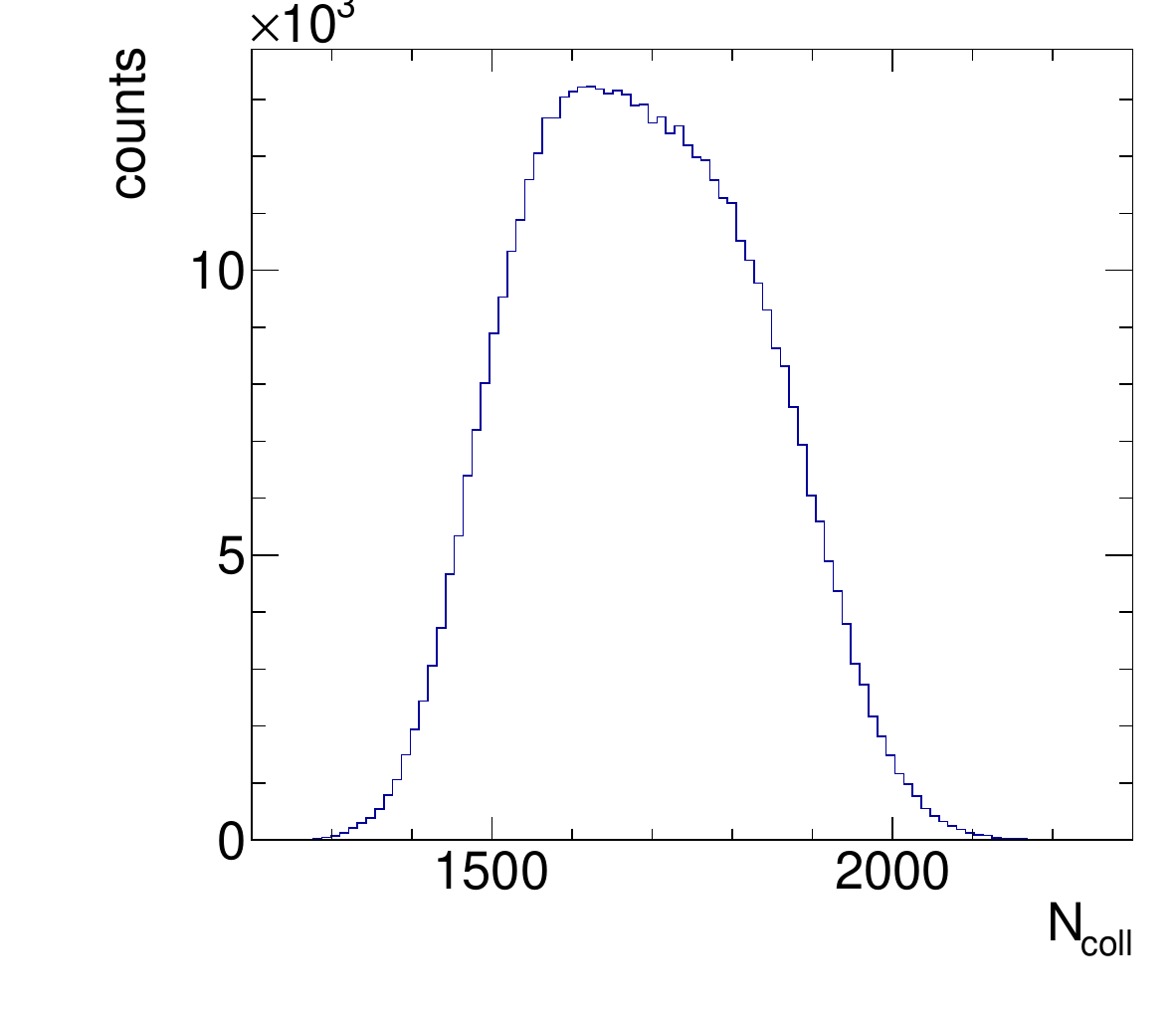}
      \caption{Number of wounded nucleons (left panel) and binary collisions (right panel) as generated with a Glauber Monte Carlo simulation using input from the ALICE experiment~\cite{ALICE:2013hur}.  The selection corresponds to the 5\% most central Pb-Pb collisions at $\sqrt{s_{\mathrm{NN}}}=2.76$~\tev.}
    \label{fig:Alice_Npart_Ncoll}
\end{figure}
\section{Model A}
In model~A we first generate the charged-particle multiplicity for individual events using the NDB distribution as extracted by experimental measurements.
In doing so we sample the NBD distribution $n_{s}$ times.
Different particle species are then taken as fractions of the total number of charged particles. 
For example, from a randomly sampled NBD distribution a respective fraction is assigned to protons.  
From the remaining charged particles another fraction is assigned to positively charged pions and the rest is taken as negatively charged pions. 
These fractions are chosen such that the overall probability of having protons, positively and negatively charged pions correspond to 75$\%$, 9$\%$ and 16 $\%$ of all charged particles, respectively, based on the actual HADES measurement in Au+Au collisions (see~\cite{Harabasz:2022rdt} and references therein).
In addition, we account for acceptance effects, because the NDB parameters are obtained in different acceptance than that used for the fluctuation analyses. 
We therefore fold the entire NDB distribution with a binomial distribution such that the experimentally measured mean multiplicities of particles in the acceptance used for fluctuation analysis are reproduced. 
Volume fluctuations are naturally accounted for as for each event the number of sources  $n_{s}$ are randomly sampled from the corresponding distributions.

The simulated mean numbers of protons are shown in the left panel of~Fig.~\ref{fig:kappa1_2Protons} for the four centrality classes. 
In the right panel of Fig.~\ref{fig:kappa1_2Protons} the reconstructed second-order cumulants of protons are presented, normalized to the expected true cumulant, $\bar{\kappa}_2[p]_{sim}$. 
The black stars represent the results which include volume fluctuations. 
The values $\kappa_{2}^{corr}$ as calculated using Eq.~\ref{eq:k2_corr_1} are shown with blue stars, while the red stars correspond to $\bar{\kappa}_{2}[p]$ = $\kappa_{2}^{corr}[p]$ + $\Delta_{2}[p]$.  The results for the third and fourth order cumulants are shown in Fig.~\ref{fig:kappa3_4Protons}. The corresponding normalized biases  $\Delta_n/\bar{\kappa}_n$ are presented in Fig.~\ref{fig:bias_model_A}. We find the normalized biases to be very small, of the order of a few present, so that the corrected cumulants, $\kappa_{n}^{corr}$ are very close to their expected true values, $\bar{\kappa}_n$. As already discussed, this is to be expected since the multiplicity distribution per wounded nucleon in the present Model is close to Poisson.

\subsection{High energy limit}
\begin{figure}[!htb]
    \centering
    \includegraphics[width=.8\linewidth,clip=true]{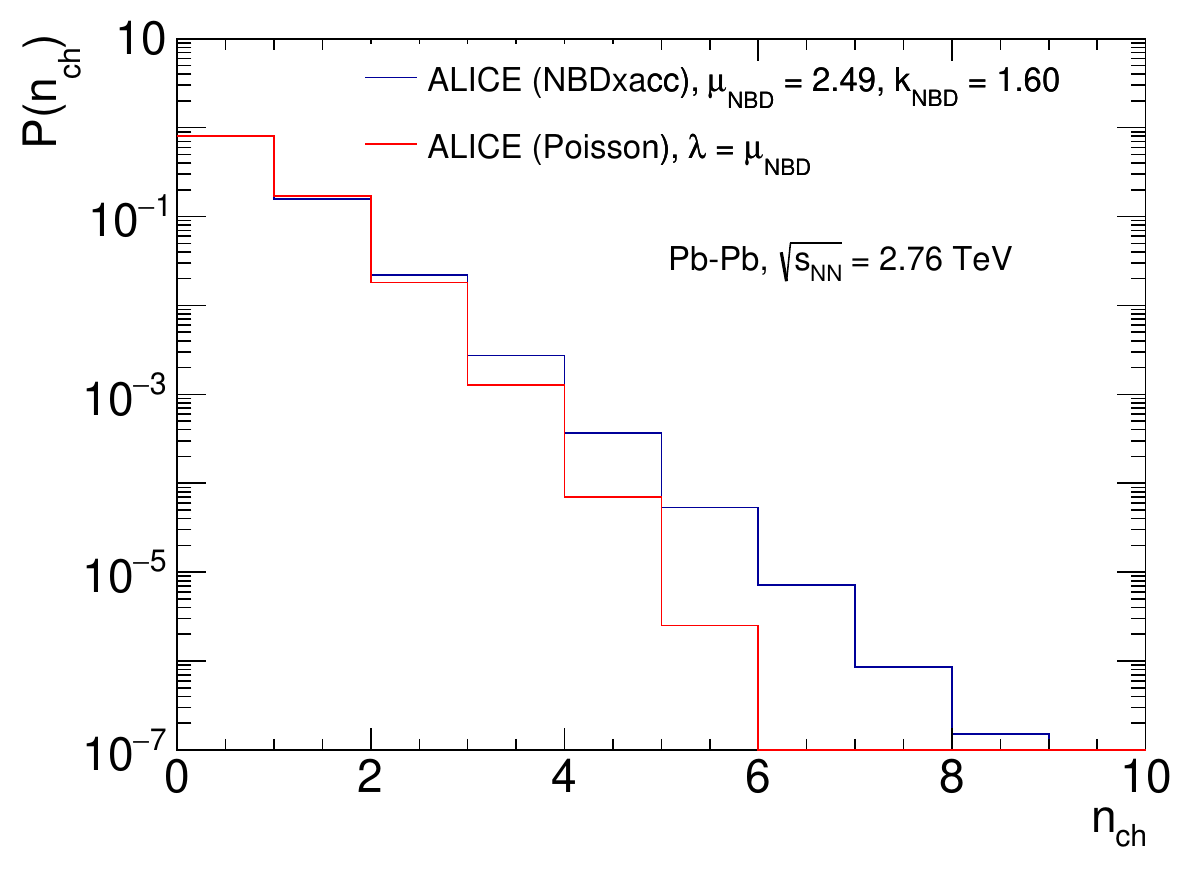}
      \caption{NDB distributions from ALICE as measured by fitting the signal amplitude in VZEROs (left panel) and after folding within the experimental acceptance 8.5$\%$ (see the text for details). The red histogram corresponds to a Poisson distribution with the same mean, $\mu$ as the folded NBD distribution. The original NBD distribution from ALICE is presented in the panel (c) of Fig.~\ref{fig:NBD_HADES_STAR}.}
    \label{fig:Alice_NDB_flded}
\end{figure}

In this section we apply the method to high energy collisions 
using ALICE data. In doing so we first run Glauber Monte Carlo simulations for Pb-Pb collisions at $\sqrt{s_{\mathrm{NN}}}=2.76$~\tev. The input parameters are taken from Ref.~\cite{ALICE:2013hur}.  
Following the ALICE experiment~\cite{ALICE:2013hur}, different centrality classes are chosen by introducing sharp cuts on the charged-particle distributions. 
The distributions of wounded nucleons and binary collisions corresponding to the 5$\%$ most central collisions are presented in Fig.~\ref{fig:Alice_Npart_Ncoll}~\cite{Braun-Munzinger:2016yjz}.
The reconstructed mean number of wounded nucleons and binary collisions, corresponding to the 5$\%$ most central collisions are $\langle N_{W}\rangle \approx 382$  and  $\langle N_{coll}\rangle \approx 1685$, respectively, consistent with the numbers given in~\cite{ALICE:2013hur}. With these numbers one can estimate a mean number of particle emitting sources, yielding $\langle n_{s}\rangle\approx 642$ (see Eq.~\ref{eqn_sources}). The corresponding mean number of charged particles can be estimated as $\langle N_{ch} \rangle = \langle n_{s}\rangle\times \mu_{NBD} \approx$ 18811. On the other hand, the total number of charged particles measured inside the ALICE acceptance is about 1601~\cite{ALICE:2013mez}. We thus folded the ALICE NBD distribution with a binomial with the acceptance factor of $\epsilon = 1601/18811 \approx 8.5\%$ to obtain the distribution within the experimental acceptance.\footnote{We note that binomial folding of the NDB distribution, with the acceptance factor $\epsilon$, changes only the parameter $\mu$ of the original NDB distribution ($\mu\rightarrow \epsilon\mu$), while the parameter $k$ stays unchanged.}  The so obtained NDB distribution from ALICE is presented in Fig.~\ref{fig:Alice_NDB_flded}. We further note that only the acceptance in rapidity is considered. Fluctuation analyses are performed within a finite momentum range. Inclusion of the latter will further reduce the discrepancy between NBD and the corresponding Poisson distribution shown with the red histogram in Fig.~\ref{fig:Alice_NDB_flded}.

Finally using the NBD distribution presented in Fig.~\ref{fig:Alice_NDB_flded}, and measured proton number, $\langle N_{p}\rangle\approx$ 35~\cite{ALICE:2013mez}, we estimated $\Delta_{2}[p]\approx$ 1.2. This corresponds to a bias  of  $\Delta_{2}[p]/\bar{\kappa}_{2}[p]\approx 3.3 \%$.

\section{Model B}
\begin{figure}[!htb]
    \centering
    \includegraphics[width=.45\linewidth,clip=true]{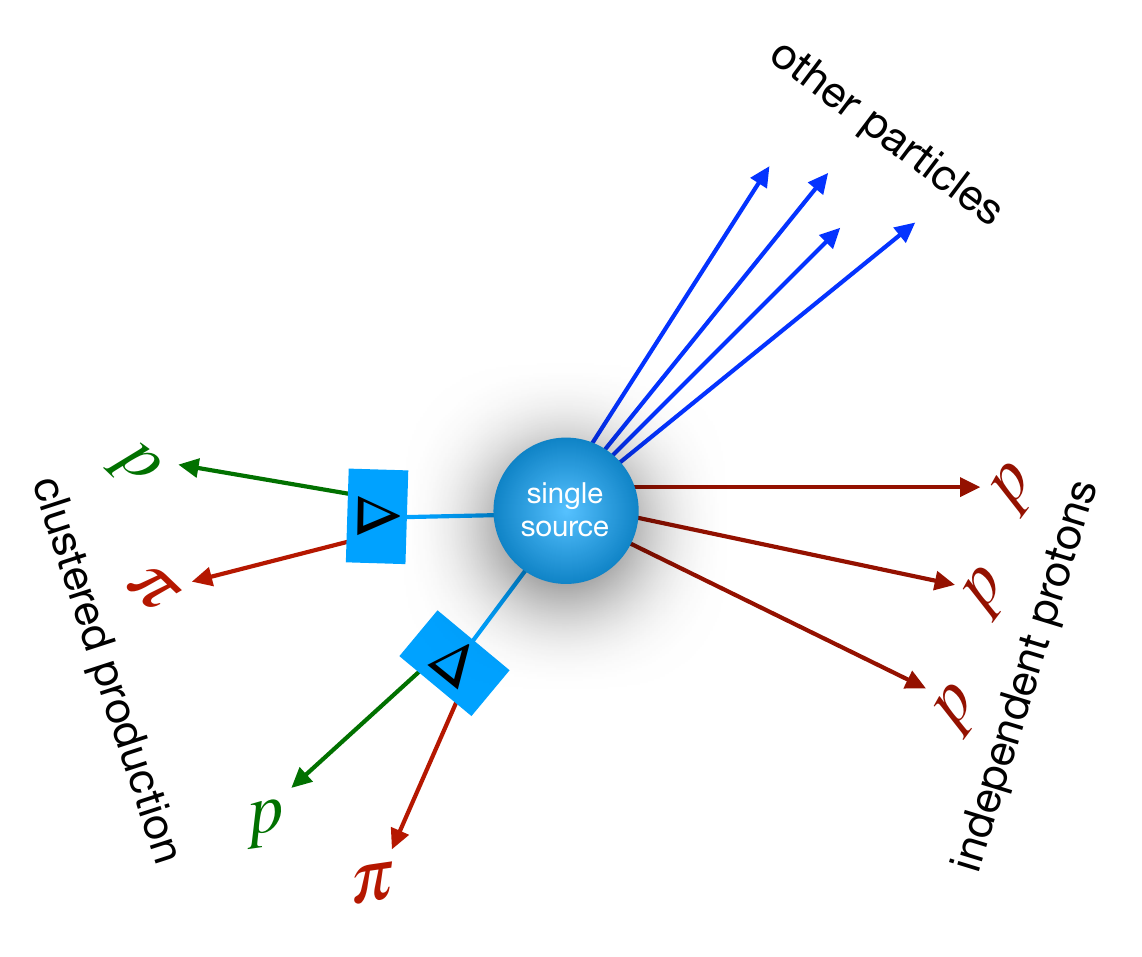}
    \includegraphics[width=.45\linewidth,clip=true]{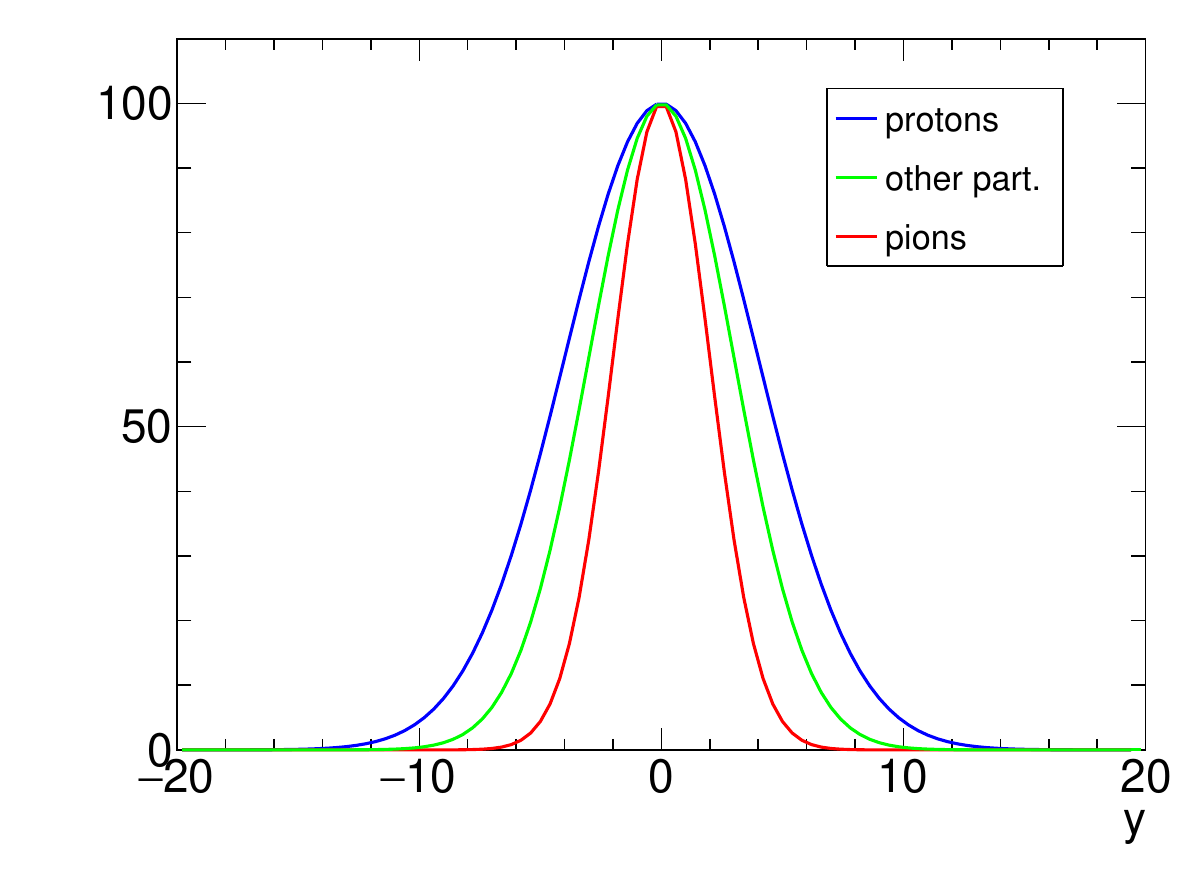}
    \caption{Left panel: Schematic illustration of model B. From a single source two clusters and three protons are produced. The clusters further decay to protons and pions. In simulations both, clusters and independent protons are sampled from  Poisson distributions. The corresponding mean values of clusters and independent protons are provided in Table~\ref{tab:modelB-parameters}. Right panel: Schematic rapidity distributions for different particles in order to study acceptance effects.}
    \label{fig:modelB}
\end{figure}

In model B we introduce correlations between charged particles, specifically pions and protons by generating clusters (or rather resonances). Especially for HADES energies most of the observed pions are believed to originate from decays of Delta resonances. Therefore, the effect of such decay correlations, while no treated quantitatively here, needs to be taken into account. Specifically, this is done by generating clusters of particles (e.g., resonances) from each source and letting them decay into two different particle species. Moreover, the clusters are generated from a Poisson distribution. In addition we produce independent particles from each source as well, sampled also from a Poisson distribution. Schematic illustration of the model for a single source is given in Fig.~\ref{fig:modelB}. Fluctuations of sources are introduced like in the model A. 
The input parameters for model B are given in Table~\ref{tab:modelB-parameters}.

\begin{table}[!htbp]
\centering
 \begin{tabular}{||c |  c ||} 
 \hline
particles & mean numbers per source   \\ [0.5ex] 
 \hline
 clusters & 0.03   \\ 
 \hline
 independent protons & 0.23  \\
 \hline
 other particles & 0.21 \\
\hline
\end{tabular}
\caption{\label{tab:modelB-parameters} Parameters for model B are mean numbers of different particles species per source (see~\cite{Harabasz:2022rdt} and references therein).  In addition, each cluster decays into one proton and one pion. Numbers of clusters, independent protons and other particles are sampled from independent Poisson distributions.}
\end{table}

By construction the simulated protons, pions and clusters follow a Poisson distribution. However the distribution of the total number of particles does not, due to the correlation between pions and protons introduced via the cluster decay (see Appendix \ref{sec:clusters}). 

In experiments measurements are performed inside the finite acceptance by imposing selection criteria in momentum space, e.g., on rapidity and/or transverse momentum of particles. Moreover, such conditions typically lead to different acceptances for different particle species. In order to study the impact of the finite acceptance on the presented formalism, we introduce arbitrary rapidity distributions for protons, pions and other particles as illustrated in the rigth panel of Fig.~\ref{fig:modelB}. To this end we generate rapidity values for protons, pions and other particles from the corresponding distributions presented in Fig.~\ref{fig:modelB}.

\begin{figure}[!htb]
    \centering
    \includegraphics[width=.45\linewidth,clip=true]{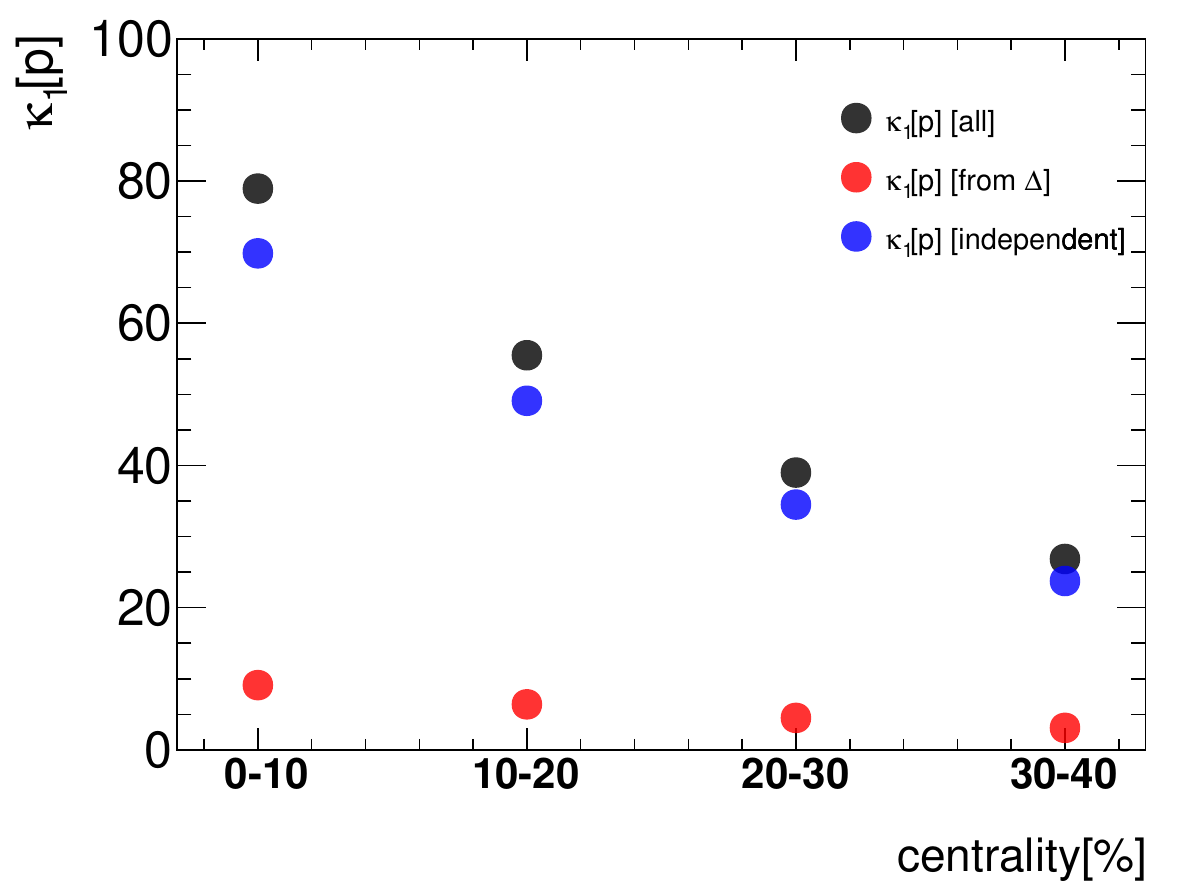}
     \includegraphics[width=.45\linewidth,clip=true]{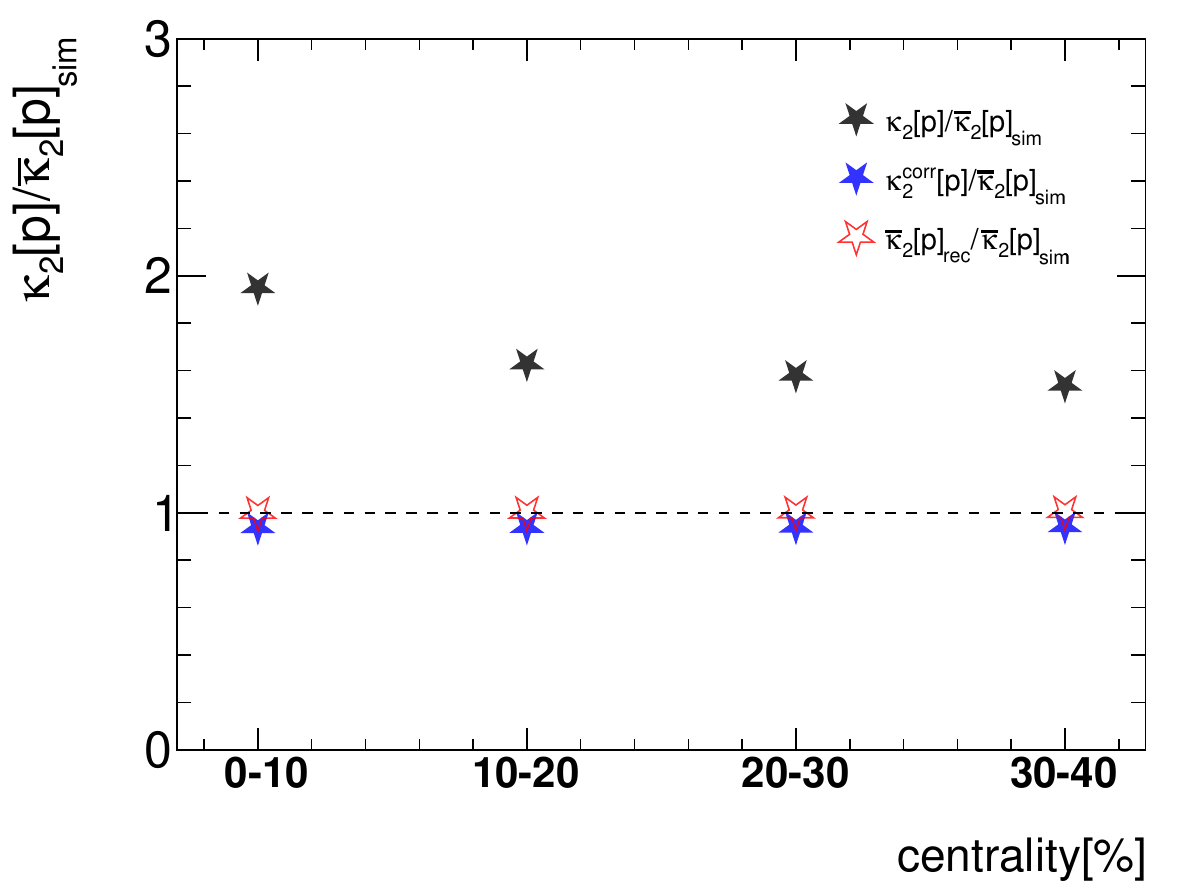}
      \caption{Left panel: Mean number of protons produced via clusters and independently are presented with the red and blue circles respectively. The black circles represent total multiplicity of protons. Pions are produced only via clusters. Right panel: Reconstructed second-order cumulants of protons including participant fluctuations (black stars). Corrected values for cumulants $\kappa_{2}^{corr}[p]$, i.e.,  without the bias term $\Delta_{2}[p]$ are presented with blue stars, while red stars represent fully corrected, against volume fluctuation. The results are normalized to the true second order cumulant, $\bar{\kappa}_{2}[p]_{sim}$.}
    \label{fig:ModelB_kappa1_kappa_2}
\end{figure}

\begin{figure}[!htb]
    \centering
    \includegraphics[width=.45\linewidth,clip=true]{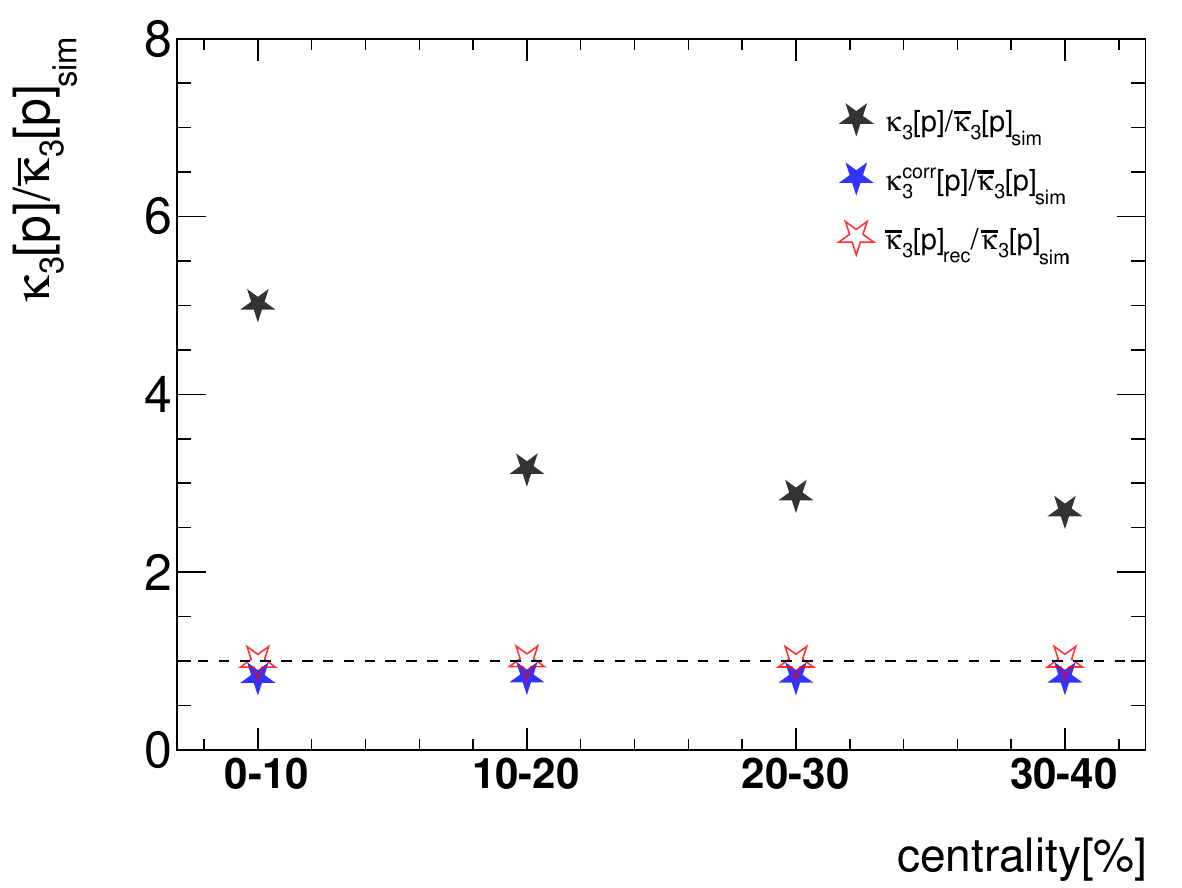}
    \includegraphics[width=.45\linewidth,clip=true]{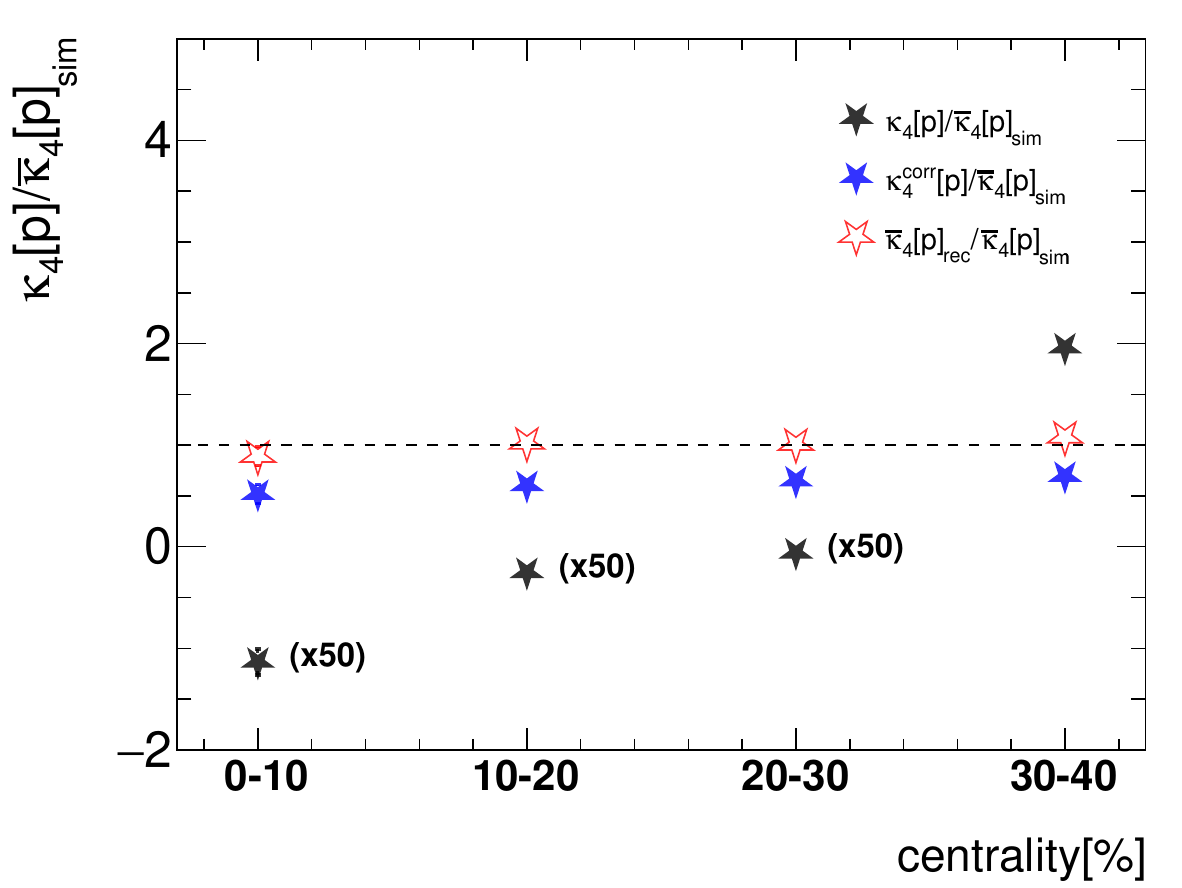}
      \caption{Left panel: Reconstructed third-order cumulants of protons including participant fluctuations (black stars) for model B. Corrected values for cumulants $\kappa_{3}^{corr}[p]$, i.e.  without the bias term $\Delta_{3}[p]$ are presented with blue stars, while red stars represent fully corrected, against volume fluctuations, values $\bar{\kappa}_{3}$[p]=$\kappa_{3}^{corr}[p]$ + $\Delta_{3}[p]$. Right panel: Similar to the left panel but for the fourth-order cumulants. Note that the values for the fourth-order cumulants with volume fluctuations (black stars) need to be multiplied by 50 for the first three centrality classes. The results are normalized to the true third or fourth order cumulant, $\bar{\kappa}_{3}[p]_{sim}$ ,$\bar{\kappa}_{4}[p]_{sim}$, respectively }
    \label{fig:ModelB_kappa3_kappa_4}
\end{figure}

\begin{figure}[!htb]
    \centering
    \includegraphics[width=.45\linewidth,clip=true]{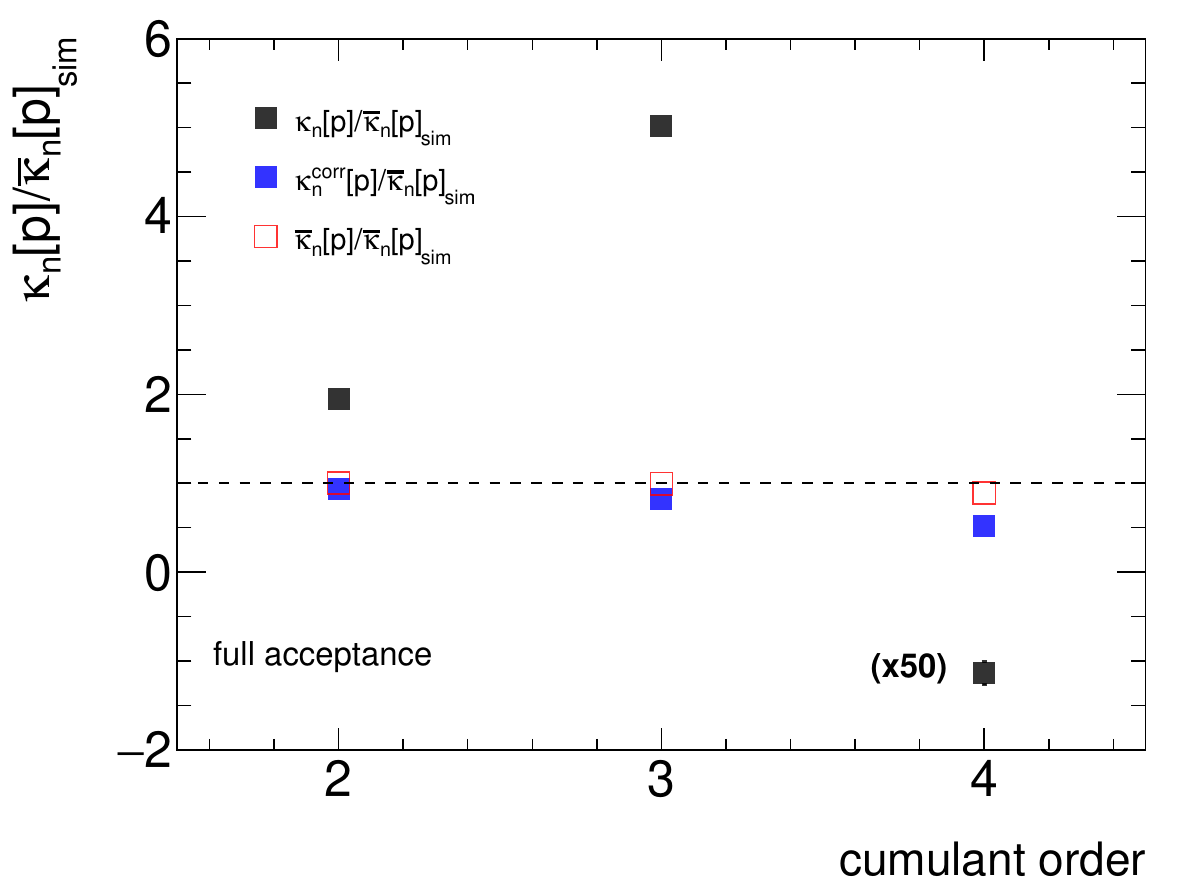}
    \includegraphics[width=.45\linewidth,clip=true]{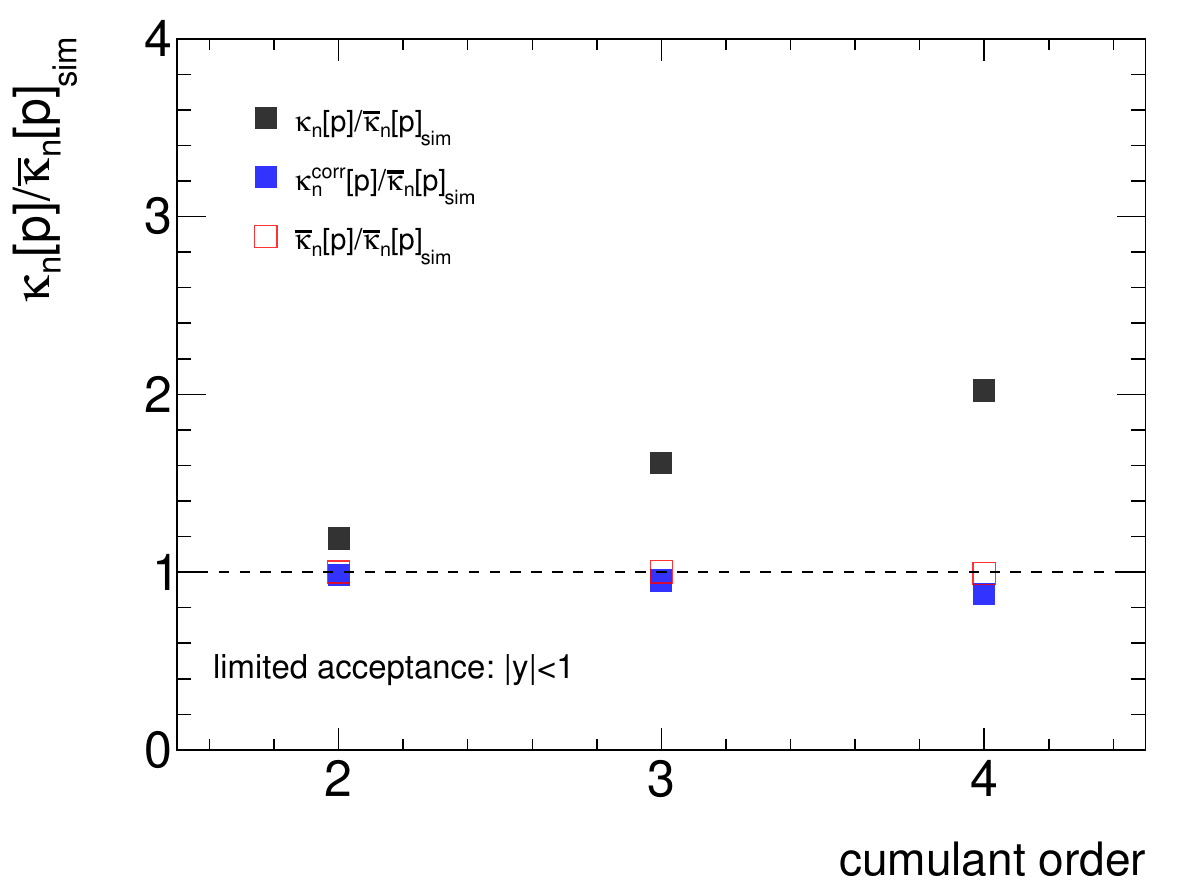}
      \caption{Left panel: Cumulants of protons in the full acceptance, presented for the 10$\%$ most central collisions (cf. Figs.~\ref{fig:ModelB_kappa1_kappa_2} and~\ref{fig:ModelB_kappa3_kappa_4}).  Right panel: Similar plot for the 10$\%$ most central collisions, but inside the finite acceptance delimited with the $|y|<1$ criterion. The results are normalized to the true cumulants, $\bar{\kappa}_{n}[p]_{sim}$.}
    \label{fig:cumsforcent0}
\end{figure}

\begin{figure}[!htb]
    \centering
    \includegraphics[width=.45\linewidth,clip=true]{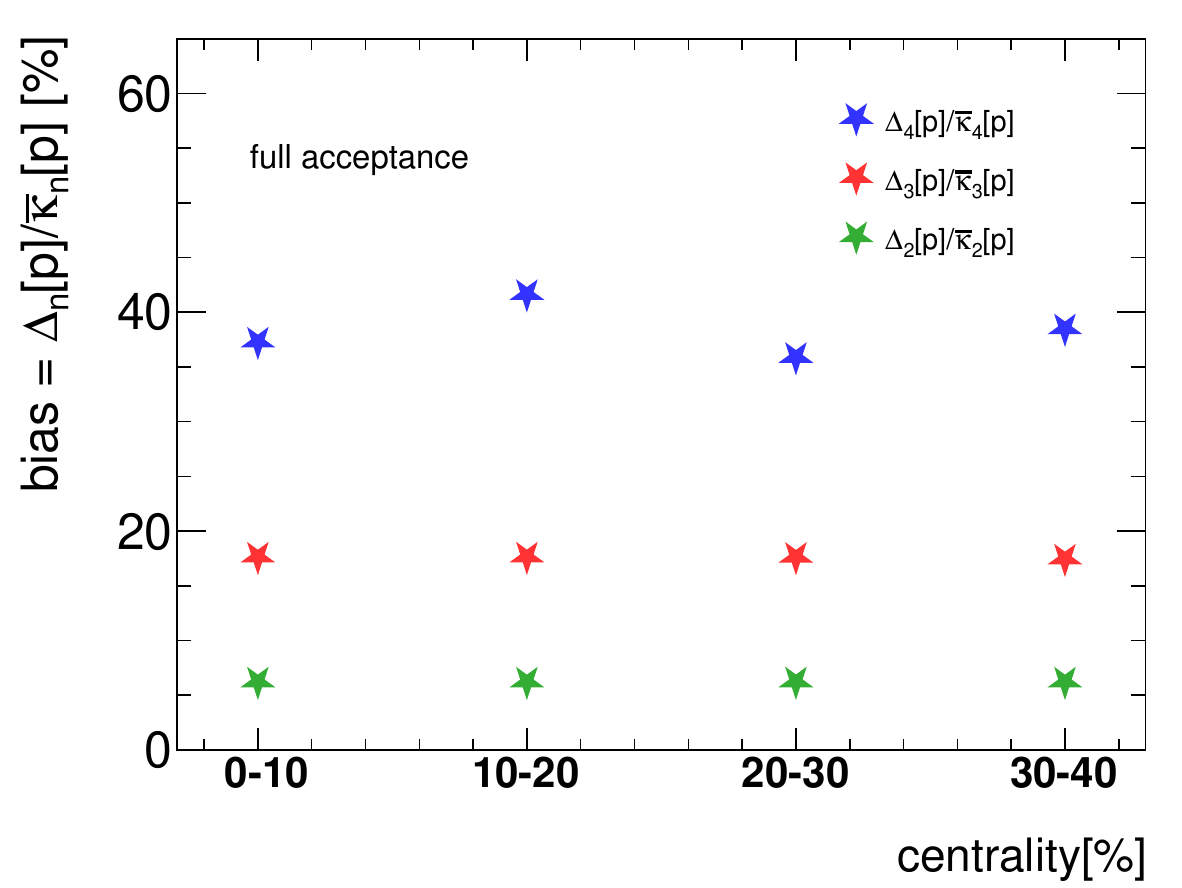}
    \includegraphics[width=.45\linewidth,clip=true]{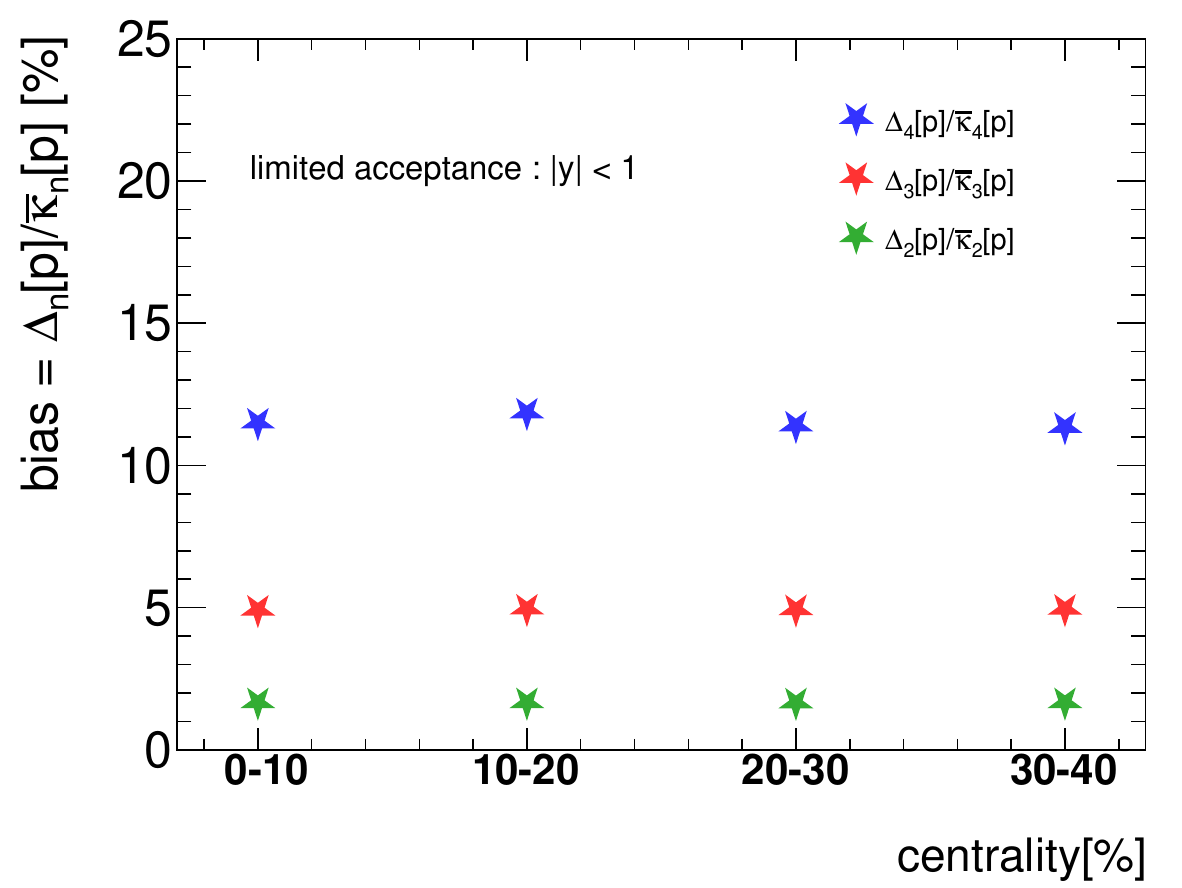}
      \caption{The bias terms for model B, in the full (left panel) and finite (right panel) acceptances. The acceptance, $|y|<1$ is introduced using  rapidity distributions of pions and protons as shown in Fig.~\ref{fig:modelB}. }
    \label{fig:ModelB_Deltas}
\end{figure}

In the left panel of Fig.~\ref{fig:ModelB_kappa1_kappa_2} we present mean multiplicities of protons produced via clusters (red circles) and independently (blue circles), while the black circles correspond to the total number of protons. In addition, we produce pions from clusters, and, by construction, their mean values are equal to those of protons from clusters. The right panel of Fig.~\ref{fig:ModelB_kappa1_kappa_2} shows the second-order cumulants of  protons  divided by the expected value  $\bar{\kappa}_2[p]_{sim}$. The black stars represent those including participant (volume)  fluctuations, $\kappa_2[p]/\bar{\kappa}_2[p]_{sim}$.  The corrected cumulants $\kappa_{2}^{corr}[p]/\bar{\kappa}_2[p]_{sim}$ (see Eq.~\ref{eq:k_corr_2}) are shown with blue symbols, while the open red stars represent the true reconstructed values of fluctuations of protons $\bar{\kappa}_{2}[p]/\bar{\kappa}_2[p]_{sim}$ as calculated using Eq.~\ref{eq_kappa_bar}. Similar results for the third and fourth order cumulants of protons are presented in Fig.~\ref{fig:ModelB_kappa3_kappa_4} (see Eqs.~\ref{eq:k_corr_3},~\ref{eq:k_corr_4},~\ref{eq_kappa_bar}).  In Fig.~\ref{fig:cumsforcent0} the normalized cumulants as a function of cumulant order are presented for the 10$\%$ most central collisions. The right panel of Fig.~\ref{fig:cumsforcent0} corresponds to the full acceptance, while in the right panel the results in the finite acceptance, delimited as $|y| < 1$, are presented. One clearly observes that in the finite acceptance the precision of the method is significantly better.   
In Fig. \ref{fig:ModelB_Deltas}  we show the magnitude of the corresponding  normalized biases, $\Delta_n/\bar{\kappa}_n$,  for the full acceptance (left panel) and for $|y|<1$ (right panel). While the biases for the full acceptance may at first sight appear rather large ($\sim 40 \%$ ) one should realize that for the most central events the uncorrected fourth order cumulant is more than a factor of 50 larger in magnitude than the true cumulants. In other words the proposed corrections, while not perfect are a huge improvement of the measurement. The situation gets better for the limited acceptance.

\section{Software package}
\begin{figure}[!htb]
    \centering
    \includegraphics[width=.9\linewidth,clip=true]{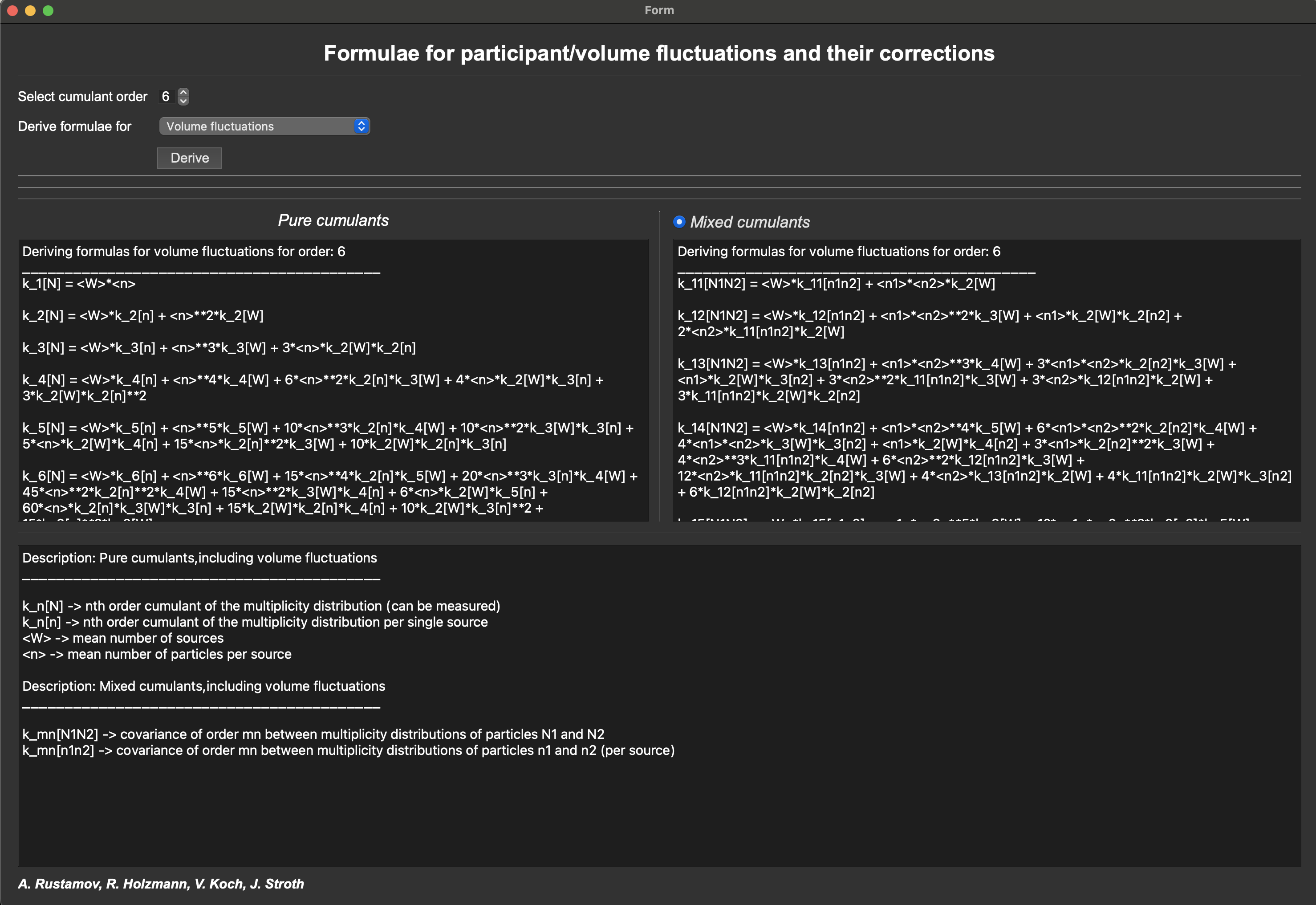}
      \caption{The GUI for deriving analytic formulas.} 
    \label{fig:GUI}
\end{figure}

A Python package is provided to derive analytical formulas for both mixed and pure cumulants of multiplicity distributions, including participant/volume fluctuations. The correction formulas and their bias terms can be derived as well. The dedicated graphical user interface is presented in Fig.~\ref{fig:GUI} and can be downloaded via Ref.~\cite{testGui1}.

\section{Discussion and Summary}
\begin{itemize}
\item We have shown that using mixed events to determine the contributions
of wounded nucleon or volume fluctuations is equivalent to extracting the latter from the track multiplicity distribution. However event mixing may offer an advantage since it allows to generate an almost arbitrarily large event ensemble with the same multiplicity distribution, and thus eliminate possible constraints due to limited event statistics.\\
In either case, not all contributions can be accessed by a direct measurement.
The remaining terms lead to biases, $\Delta_{k}$, which depend
on the multiplicity distribution per wounded nucleon. These biases
are, however, parametrically suppressed  by powers of $\ave N/\ave M$ depending on the order of the cumulants. The biases are also small if the multiplicity distribution per wounded nucleon is close to Poisson.
In addition,  we suggest to constrain these biases in experiment with fits to the observed multiplicity distribution within the wounded-nucleon model.
\item We have worked here within the wounded nucleon model to formulate 
volume fluctuations. Alternatively, one may introduce generic volume
fluctuations as done e.g.\ in \cite{Skokov:2012ds}. It is easy to show
(see Appendix \ref{sec:VolumeFluctuations}) that the resulting expressions for the corrected cumulants, $\kappa_{j}^{corr}$, and the biases, $\Delta_{j}$, are identical to those derived here, i.e.\ Eqs. (\ref{eq:k_corr_2}-\ref{eq:k_corr_4}). 
\item We note that one gets similar expressions for the fluctuations from the wounded nucleons, Eqs. (\ref{eq:k_wound_2}-\ref{eq:k_wound_4}), in terms of the factorial cumulants of, for example, pions instead of the total track multiplicity. This has the advantage that the corrections do not involve the particles of interest, protons, in our case.  Of course, this approach requires that sufficiently many pions are produced and thus may be limited to collisions at higher energies.
\item  We have checked that the proposed method also works if the multiplicity distribution is determined for a different acceptance than the particle distribution of interest. In this case all quantities in the expression for the corrected cumulants, Eqs.(\ref{eq:k_corr_2}-\ref{eq:k_corr_4}) involving the multiplicity should be evaluated in the multiplicity acceptance while all quantities involving the particles of interest, such as $\ave N$ or $\kappa_j[N]$ should be determined in the particle acceptance.
\item We have verified that the proposed method is not affected by potentially different rapidity distributions for different particle types.
\item We note that the correction term and bias for the second-order cumulant depends on the properties of the multiplicity distribution only while those for higher-order cumulants also involve the (uncorrected) cumulants of interest (at a lower order), $\kappa_n[N]$ (see Eq. (\ref{eq:k_corr_3},\ref{eq:k_corr_4}).  
\item The proposed method is also applicable for mixed cumulants, such as the covariance between two particle species. The relevant formulas for mixed cumulants between species $A$ and $B$ up to $\kappa_{2,2}[A,B]$ are given in Appendix \ref{sec:mixed_cumulants}. 

The corresponding relations for correction and bias terms for mixed cumulants of any order can be obtained with the provided software package~\cite{testGui1}.

\item Here we  have not explicitly discussed corrections for net proton cumulants. However, those can be easily obtained using the corrections to the mixed cumulants provided in Appendix \ref{sec:mixed_cumulants}. For example, the second-order cumulant of the net proton distribution is given by
\begin{align}
    \kappa_2[N-\bar{N}] = \kappa_2[N]+\kappa_2[\bar{N}]- 2 \kappa_{1,1}[N,\bar{N}]
\end{align} 
The corrected cumulant is the (using Eqs. (\ref{eq:k_corr_2}, \ref{eq:kappa_11}) 
\begin{align}
    \kappa_2^{corr}[N-\bar{N}]=\kappa_2[N-\bar{N]} -\frac{\left( \ave{N} - \ave{\bar{N}}\right)^2}{\ave{M}^2}C_2[M]
\end{align}
with the bias
\begin{align}
    \Delta = \frac{\left( \ave{N} - \ave{\bar{N}}\right)^2}{\ave{M}^2}\bar{C}_2[M]
\end{align}
For systems at vanishing baryon number chemical potential, such as those created at very high collision energies, $\ave{N}=\ave{\bar{N}}$ so that the corrected cumulant is identical to the measured one (as discussed in \cite{Skokov:2012ds} ) and the bias vanishes.

\end{itemize}

In summary, we have presented a method to correct experimentally measured particle number cumulants for the effect of participant or volume fluctuations. The essential idea is to extract the contribution from the volume fluctuations from the distribution of charged particles which, after appropriate re-scaling, may be subtracted from the measured cumulants of interest. Our proposed method is not exact as there remains a bias or remnant which can not be accessed directly  from experiment. However, we have shown by model calculations that this bias is very small compared to the contribution from participant fluctuations and we hence consider our method an important step towards measuring the true dynamical fluctuations of the system.  

\begin{acknowledgments}
V.K would like to thank GSI and the Institute for Nuclear Theory at the University of Washington for their kind hospitality and stimulating research environment. V.K. has been supported by the U.S. Department of Energy, Office of Science, Office of Nuclear Physics, under contract number DE-AC02-05CH11231, by the INT's U.S. Department of Energy grant No. DE-FG02-00ER41132, and by the ExtreMe Matter Institute EMMI at the GSI Helmholtzzentrum f{\"u}r Schwerionenforschung, Darmstadt, Germany.

\end{acknowledgments}

\appendix

\section{Wounded-nucleon model\label{sec:Wounded-nucleon-model}}

Here we briefly discuss the wounded-nucleon model following the Appendix
of Ref. \citep{Bzdak:2016jxo}. The wounded-nucleon model assumes
that particles are produced by independent sources, called wounded
nucleons or participants. Therefore, the probability to find $A$
particles of type $A$ and $B$ particles of type $B$ can be written
as 
\begin{equation}
P\left(A,B\right)=\sum_{w}W\left(w\right)\sum_{a_{1},\cdots a_{w}}\sum_{b_{1},\cdots b_{w}}p\left(a_{1},b_{1}\right)\cdots p\left(a_{w},b_{w}\right)\,\delta_{A,\sum_{k=1}^{w}a_{k}}\delta_{B,\sum_{k=1}^{w}b_{k}}.\label{eq:real_probability}
\end{equation}
Here $W(w)$ denotes the probability to have $w$ wounded nucleons,
and $p(a,b)$ is the probability to have $a$ particles of type $A$
and $b$ particles of type $B$ from one wounded nucleon. The moment-generating function, $h\left(t_{A},t_{B}\right)$ is then
\begin{align}
H\left(t_{A},t_{B}\right) & =\sum_{A,B}e^{t_{A}A}e^{t_{b}B}P\left(A,B\right)\nonumber \\
 & =\sum_{w}W\left(w\right)\sum_{a_{1},\cdots a_{w}}\sum_{b_{1},\cdots b_{w}}p\left(a_{1},b_{1}\right)\cdots p\left(a_{w},b_{w}\right)e^{t_{A}\sum_{k=1}^{w}a_{k}}e^{t_{B}\sum_{k=1}^{w}b_{k}}\nonumber \\
 & =\sum_{w}W\left(w\right)\sum_{a_{1},b_{1}}p\left(a_{1},b_{1}\right)e^{t_{A}a_{1}+t_{b}b_{1}}\cdots\sum_{a_{w},b_{w}}p\left(a_{w},b_{w}\right)e^{t_{A}a_{w}+t_{B}b_{w}}\nonumber \\
 & =\sum_{w}W\left(w\right)\left[\sum_{a,b}p\left(a,b\right)e^{t_{A}a+t_{B}b}\right]^{w}\nonumber \\
 & =\sum_{w}W\left(w\right)\left[h_{w}\left(t_{A},t_{B}\right)\right]^{w}\nonumber \\
 & =\sum_{w}W\left(w\right)e^{w\,g_{w}(t_{A},t_{B})}\label{eq:moment_gen}
\end{align}
where $h_{w}\left(t_{A},t_{B}\right)=\sum_{a,b}p\left(a,b\right)e^{t_{A}a+t_{B}b}$
is the moment generating function and $g_{w}(t_{A},t_{B})=\ln\left[h_{w}\left(t_{A},t_{B}\right)\right]$
the cumulant-generating function for one wounded nucleon, respectively.
The cumulant-generating function, $G\left(t_{A},t_{B}\right)=\ln\left(h\left(t_{A},t_{B}\right)\right)$,
is then given by

\begin{equation}
G\left(t_{A},t_{B}\right)=\ln\left[H\left(t_{A},t_{B}\right)\right]=\ln\left[\sum_{w}W\left(w\right)e^{w\,g_{w}(t_{1},t_{2})}\right]=G_{W}\left(g_{w}(t_{A},t_{B})\right)\label{eq:cum_gen_wound}
\end{equation}
where $G_{W}(t)=\ln\left[\sum_{w}W\left(w\right)e^{w\,t}\right]$
in the cumulant-generating function for the wounded-nucleon distribution,
$W(w)$. We note, that $g_{w}(0,0)=G_{W}(0)=0$ by construction. The
cumulants are then obtained as
\begin{equation}
  \kappa_{j,k}[A,B]=\left.\frac{\partial^{j}\partial^{k}}{\partial t_{A}\partial t_{B}}g\left(t_{A},t_{B}\right)\right|_{t_{A}=t_{B}=0}.
\label{eq:mixed_cum_wound}
\end{equation}
For example:

\begin{align}
\kappa_{1}[A] & =\frac{\partial}{\partial t_{A}}\left.G(t_{A},0)\right|_{t_{A}=0}=\frac{dG_{w}}{dg_{w}}\left.\frac{dg_{w}(t_{A},0)}{dt_{A}}\right|_{t_{A}=0}=\left.\frac{dG_{w}}{dg_{w}}\right|_{g_{w}=0}\left.\frac{dg_{w}(t_{A},0)}{dt_{A}}\right|_{t_{A}=0}\nonumber \\
 & =\kappa_{1}[N_{w}]\kappa_{1}[a]=\ave{N_{w}}\ave a\label{eq:k1_wound}
\end{align}
where $\kappa_{1}[n]=\ave a$ denotes the mean number of particles
of type $A$ per wounded nucleon and $\kappa_{1}[w]=\ave{N_{w}}$
the mean number of wounded nucleons. The second-order cumulant is
\begin{align}
\kappa_{2}[A] & =\frac{\partial^{2}}{\partial t_{A}^{2}}\left.G(t_{A},0)\right|_{t_{A}=0}\nonumber \\
 & =\frac{d^{2}G_{w}}{dg_{w}^{2}}\left.\left(\frac{dg_{w}(t_{A},0)}{dt_{A}}\right)^{2}\right|_{t_{A}=0}+\frac{dG_{w}}{dg_{w}}\left.\frac{d^{2}g_{w}(t_{A},0)}{dt_{A}^{2}}\right|_{t_{1}=0}\nonumber \\
 & =\kappa_{2}[N_{w}]\kappa_{1}[a]^{2}+\kappa_{1}[N_{w}]\kappa_{2}[a]=\kappa_{2}[N_{w}]\ave a^{2}+\ave{N_{w}}\kappa_{2}[a]\label{eq:k2_wound}
\end{align}
The covariance is
\begin{align}
cov[A,B] & =\frac{\partial}{\partial t_{A}\partial t_{B}}\left.G(t_{A},t_{B})\right|_{t_{A},t_{B}=0}=\frac{\partial}{\partial t_{B}}\left.\left(\frac{dG_{w}}{dg_{w}}\frac{\partial g_{w}(t_{A},t_{B})}{\partial t_{A}}\right)\right|_{t_{A},t_{B}=0}\nonumber \\
 & =\left.\frac{d^{2}G_{w}}{dg_{w}^{2}}\frac{\partial g_{w}(t_{A},t_{B})}{\partial t_{A}}\frac{\partial g_{w}(t_{A},t_{B})}{\partial t_{B}}\right|_{t_{A},t_{B}=0}+\left.\left(\frac{dG_{w}}{dg_{w}}\frac{\partial^{2}g_{w}(t_{A},t_{B})}{\partial t_{A}\partial t_{B}}\right)\right|_{t_{A},t_{B}=0}\nonumber \\
 & =\kappa_{2}[N_{W}]\kappa_{1}[a]\kappa_{1}[a]+\kappa_{1}[N_{w}]cov[a,b]\nonumber \\
 & =\kappa_{2}[N_{w}]\ave a\ave b+\ave{N_{w}}cov[a,b]\label{eq:cov_wound}
\end{align}

Using the relation between the cumulant and factorial cumulant generating
function, Eq. \ref{eq:fac_cum_gen_from_cum_gen} , the factorial cumulant
generating function is given by
\begin{equation}
G_{F}\left(z_{\rm A},z_{\rm B}\right)=G\left(\ln(z_{\rm A}),\ln(z_{\rm B})\right)=G_{W}\left(g_{w}\left(\ln(z_{\rm A}),\ln(z_{\rm B})\right)\right)=G_{W}\left(g_{F,w}\left(z_{\rm A},z_{\rm B}\right)\right),\label{eq:fac_cum_gen_wound}
\end{equation}
with 
\[
g_{F,w}\left(z_{\rm A},z_{\rm B}\right)=\sum_{a,b}p\left(a,b\right)z_{\rm A}^{a}z_{\rm B}^{b}
\]
the factorial cumulant generating function for the distribution of
one wounded nucleon, $p\left(a,b\right)$. The structure is the same
as for the cumulant generating function, except that now the argument
of the wounded-nucleon cumulant-generating function is the factorial
cumulant generating function, $g_{F,w}$. Thus the factorial cumulants
are easily obtained by simply replacing all the cumulants of the particle
distribution for a given wounded nucleon, $\kappa_{i,j}[a,b]$ with
the corresponding factorial cumulants, with $C_{1}[X]=\kappa_{1}[X]=\ave X$
\begin{align}
C_{1}[A] & =\frac{\partial}{\partial z_{\rm A}}\left.g_{F}(z_{\rm A},1)\right|_{z_{\rm A}=1}=\kappa_{1}[N_{w}]C_{1}[a]=\ave{N_{w}}\ave a\nonumber \\
C_{2}[A] & =\kappa_{2}[N_{w}]\ave a^{2}+\ave{N_{w}}C_{2}[a]\nonumber \\
C_{1,1}[A,B] & =cov[A,B]\label{eq:fac_cum_wound}
\end{align}

\section{Cumulant and factorial cumulant generating functions\label{sec:Cumulants-and-factorial} }

Given a multiplicity distribution for particles of type $A$ and $B$,
$P\left(A,B\right)$ the generating functions for cumulants, $g(t_{A},t_{B})$,
and factorial cumulants. $g_{F}(t_{A},t_{B})$ are given by
\begin{align}
g\left(t_{A},t_{B}\right) & =\ln[\sum_{A,B}P\left(A,B\right)e^{t_{A}A}e^{t_{B}B}\label{eq:cum_gen}\\
g_{F}\left(z_{\rm A},z_{\rm B}\right) & =\ln[\sum_{A,B}P\left(A,B\right)(z_{\rm A})^{A}(z_{\rm B})^{B}.\label{eq:fac_cum_gen}
\end{align}
By construction, $g\left(t_{A}=0,t_{B}=0\right)=0$ and $g_{F}\left(z_{\rm A}=1,z_{\rm B}=1\right)=0$.
Cumulants of order $(i,j)$, $\kappa_{j,k}[A,B],$ are then obtained
through 
\[
\kappa_{j,k}[A,B]=\left.\frac{\partial^{j}\partial^{k}}{\partial t_{A}\partial t_{B}}g\left(t_{A},t_{B}\right)\right|_{t_{A}=t_{B}=0},
\]
while the factorial cumulants, $C_{j,k}[A,B]$, are given by
\[
C_{j,k}[A,B]=\left.\frac{\partial^{j}\partial^{k}}{\partial z_{\rm A}\partial z_{\rm B}}g\left(z_{\rm A},zt\right)\right|_{z_{\rm A}=z_{\rm B}=1}.
\]
The generating functions are related through
\begin{equation}
g_{F}\left(z_{\rm A},z_{\rm B}\right)=g\left[\ln(z_{\rm A}),\ln(z_{\rm B})\right]\label{eq:fac_cum_gen_from_cum_gen}
\end{equation}
or vice versa
\begin{equation}
g\left(t_{A},t_{B}\right)=g_{F}\left(e^{t_{B}},e^{t_{B}}\right)\label{eq:cum_gen_fac_cum_gen}
\end{equation}

These relations may also be used to convert cumulants into factorial
cumulants and vice versa. For example, for the diagonal cumulants,
$\kappa_{n}[A]$ we have 

\begin{equation}
\kappa_{n}[A]=\sum_{j=1}^{n}S\left(n,j\right)C_{j}[A]
\end{equation}
where $S(n,j)$ denotes the Sterling numbers of the second kind. The
inverse relation is 

\begin{equation}
C_{n}[A]=\sum_{j=1}^{n}B_{n,j}\left(1,-1,2,\ldots,(-1)^{j-1}(n-j+1)!\right)\,\kappa_{j}[A]\label{eq:fac_cum_fom_cum_Bell}
\end{equation}
 with $B_{n,j}$ being Bell polynomials. For the first four orders
this evaluates to
\begin{align}
\kappa_{1} & =C_{1}\nonumber \\
\kappa_{2} & =C_{1}+C_{2}\nonumber \\
\kappa_{3} & =C_{1}+3C_{2}+C_{3}\nonumber \\
\kappa_{4} & =C_{1}+7C_{2}+6C_{3}+C_{4}\label{eq:cum_from_fac_cum}
\end{align}
and
\begin{align}
C_{2} & =\kappa_{2}-\kappa_{1}\nonumber \\
C_{3} & =2\kappa_{1}-3\kappa_{2}+\kappa_{3}\nonumber \\
C_{4} & =-6\kappa_{1}+11\kappa_{2}-6\kappa_{3}+\kappa_{4}\label{eq:fac_cum_from_cum}
\end{align}

\section{Multiplicity Distribution\label{sec:Multiplicity-Distribution}}

The multiplicity distribution, $P(M)$, is given by summing
over all (charged) particles, 
\begin{align}
P_{M}\left(M\right) & =\sum_{A,B,X}P\left(A,B,X\right)\delta_{M,A+B+X},
\end{align}
 where 
\[
P\left(A,B,X\right)=\sum_{w}W\left(w\right)\sum_{a_{1},\cdots a_{w}}\sum_{b_{1},\cdots b_{w}}\sum_{x_{1}.\cdots x_{w}}p\left(a_{1},b_{1},x_{1}\right)\cdots p\left(a_{w},b_{w},x_{w}\right)\,\delta_{A,\sum_{k=1}^{w}a_{k}}\delta_{B,\sum_{k=1}^{w}a_{k}}\delta_{X,\sum_{k=1}^{w}x_{k}}
\]
is the distribution of particles of type $A$, $B$ and all others,
denoted by $X$. The distribution for particles $A$ and $B$ defined
in Appendix \ref{sec:Wounded-nucleon-model} are then given by $P\left(A,B\right)=\sum_{X=0}^{\infty}P\left(A,B,X\right)$
while that for the particles per wounded nucleons are given by $p\left(a,b\right)=\sum_{x=0}^{\infty}p\left(a,b,x\right)$

The moment generating function is then given by (proceeding analogously
to Eq. \ref{eq:moment_gen}:
\begin{align*}
H_{M}(t) & =\sum_{M}P_{M}(M)\,e^{tM}=\sum_{M,A,B,X}P(A,B,X)\delta_{M,A+B+X}\,e^{tM}=\sum_{A,B,X}P(A,B,X)\,e^{t(A+B+X)}\\
 & =\sum_{w}W\left(w\right)\sum_{a_{1},\cdots a_{w}}\sum_{b_{1},\cdots b_{w}}\sum_{x_{1}.\cdots x_{w}}p\left(a_{1},b_{1},x_{1}\right)\cdots p\left(a_{w},b_{w},x_{w}\right)e^{t\sum_{k=1}^{w}a_{k}}e^{t\sum_{k=1}^{w}b_{k}}e^{t\sum_{k=1}^{w}x_{k}}\\
 & =\sum_{w}W\left(w\right)\sum_{a_{1}}\sum_{b_{1}}\sum_{x_{1}}p\left(a_{1},b_{1},x_{1}\right)e^{t(a_{1}+b_{1}+c_{1})}\cdots\sum_{a_{w}}\sum_{b_{w}}\sum_{x_{w}}p\left(a_{w},b_{w},x_{w}\right)e^{t(a_{w}+b_{w}+c_{w})}\\
 & =\sum_{w}W\left(w\right)\left[\sum_{a}\sum_{b}\sum_{x}p\left(a,b,x\right)e^{t(a+b+c)}\right]^{w}\\
 & =\sum_{w}W\left(w\right)\left[h_{m,w}(t)\right]^{w}\\
 & =\sum_{w}W\left(w\right)e^{wg_{m,w}(t)}
\end{align*}
where, $h_{m,w}(t)$ the moment generating function and $g_{m,w}(t)=\ln\left[h_{m,w}(t)\right]$
the cumulant generating function of the multiplicity distribution
for \emph{one} wounded nucleon, $p(m)=\sum_{a,b,x}p(a,b,x)\delta_{m,a+b+c}$.
The cumulant generating function, $G_{M}(t)$, for the multiplicity
distribution, $P\left(M\right)$, is then given by 
\begin{equation}
G_{M}(t)=\ln\left[H_{M}(t)\right]=\ln\left[\sum_{w}W\left(w\right)e^{wg_{m,w}(t)}\right]=G_{W}\left(g_{m,w}(t)\right)\label{eq:cum_gen_mult_wound}
\end{equation}
 The cumulants of the multiplicity distribution are given by
(following the analogous Eqs. \ref{eq:k1_wound} and \ref{eq:k2_wound})
\begin{align}
\kappa_{1}[M] & =\ave{N_{w}}\ave m\nonumber \\
\kappa_{2}[M] & =\kappa_{2}[N_{w}]\ave m^{2}+\ave{N_{w}}\kappa_{2}[m]\label{eq:mult_cum_wound}
\end{align}
where $\kappa_{i}[m]$ denote the cumulants of the multiplicity distribution
of \emph{one }wounded nucleon and $\kappa_{i}\left[N_{W}\right]$
those of the wounded nucleon distribution. Analogous to Eq. \ref{eq:fac_cum_gen_wound}
the factorial cumulant generating function is readily obtained 
\begin{equation}
G_{F,M}\left(z,\right)=G_{M}\left(\ln(z)\right)=G_{W}\left(g_{m,w}\left(\ln(z)\right)\right)=G_{W}\left(g_{F,m,w}\left(z\right)\right),\label{eq:fac_cum_gen_mult_wound}
\end{equation}
with 
\[
g_{F,m,w}(z)=\sum_{m}p(m)z^{m}
\]
the factorial cumulant generating function for the multiplicity distribution
of one nucleon, $p(m)$. Again, the factorial cumulants are obtained
by replacing the cumulants of the distribution $p(m)$, $\kappa_{i}[m]$
with the corresponding factorial cumulants, $C_{i}[m]$, in Eq. \ref{eq:mult_cum_wound}
by the factorial cumulants
\begin{align}
C_{1}[M] & =\ave{N_{w}}\ave m\nonumber \\
C_{2}[M] & =\kappa_{2}[N_{w}]\ave m^{2}+\ave{N_{w}}C_{2}[m]\label{eq:mult_fac_cum_wound}
\end{align}

\section{Wounded Nucleon vs Volume Fluctuations\label{sec:VolumeFluctuations}}

Here we will discuss the relation between wounded nucleon fluctuations
and so-called volume fluctuations as they are discussed e.g. in \cite{Skokov:2012ds}.
Following Ref. \cite{Skokov:2012ds} the cumulant generating function
is given by
\begin{equation}
\Phi(t)=\ln\text{\ensuremath{\left[\int dVP(V)e^{V\xi(t)}\right]}=\ensuremath{\chi^{V}(\xi(t))}}\label{eq:vol_gen_func}
\end{equation}
with $\chi^{V}(t)$ the cumulant generating function for the distribution
of volumes, $P(V)$, and 
\begin{equation}
\xi(t)=\frac{1}{V}\ln\left[\sum_{N}p(N;V)e^{Nt}\right]\label{eq:gen_kum_per_vol}
\end{equation}
the generating function for scaled cumulants, $\kappa/V$, given for
the distribution of particles at fixed volume, $p(N;V)$. Then, for
a fixed volume $V$, the scaled cumulants are given by 
\begin{equation}
\frac{\kappa_{j}}{V}=\frac{\partial^{j}}{\partial\,t^{j}}\left.\xi(t)\right|_{t=0}.\label{eq:cum_vol}
\end{equation}
For the wounded nucleon model we have (see Appendix \ref{sec:Wounded-nucleon-model})

\begin{equation}
G(t)=\ln\left[\sum_{w}W\left(w\right)e^{w\,g_{w}(t)}\right]=G_{W}(g_{w}(t))\label{eq:wound_gen_func}
\end{equation}
with $G_{W}(t)$ the cumulant generating function for the wounded
nucleon distribution, $W(w),$ and 
\begin{equation}
g_{w}(t)=\ln\left[\sum_{n}p(n)e^{nt}\right]\label{eq:gen_cum_per_wound}
\end{equation}
the generating function for the distribution of particles for one
wounded nucleon. The cumulants per wounded nucleons for a fixed number
of wounded nucleons, $N_{w},$ are given by

\begin{equation}
\frac{\kappa_{j}[N]}{N_{w}}=\kappa_{j}[n]=\frac{\partial^{j}}{\partial\,t^{j}}\left.g_{w}(t)\right|_{t=0}.\label{eq:cum_wound}
\end{equation}
Comparing the above expressions, one finds that the cumulants for volume
fluctuations can be obtained from those for the wounded-nucleon number by the
following replacements
\begin{align*}
\kappa_{j}[N_{w}] & \rightarrow\kappa_{j}[V]\\
\Kb j[N]=\ave{N_{w}}\kappa_{j}[N] & \rightarrow\ave V\frac{\kappa_{j}}{V}
\end{align*}
Indeed comparing the second-order cumulants for both scenarios we
have
\begin{align*}
\kappa_{2}[N] & =\ave{N_{w}}\kappa_{2}[n]+\ave n^{2}\kappa_{2}[N_{w}]=\Kb 2[N]+\ave N^{2}\frac{\kappa_{2}[N_{w}]}{\ave{N_{w}}^{2}}\\
\kappa_{2}[N] & =\ave V\frac{\kappa_{2}}{V}+\left(\frac{\kappa_{1}}{V}\right)^{2}\kappa_{2}[V]=\Kb 2[N]+\ave N^{2}\frac{\kappa_{2}[V]}{\ave V^{2}}
\end{align*}
where in the second line we used $\kappa_{1}=\ave N$ and $\Kb 2=\ave V\frac{\kappa_{2}}{V}$.
Obviously, analogous replacements also hold for the factorial cumulants
\[
\Cb j[N]=\ave{N_{w}}C_{j}[n]\rightarrow\ave V\frac{C_{j}}{V}
\]
with $C_{j}/V$ the volume scaled factorial cumulants.

\section{Results for  factorial cumulants}
\label{sec:results_factorial_cumulants}

Here we provide the formulas for the corrected factorial cumulants,
$C_{k}^{corr}$ and $c_{k}^{corr}$, and the the associated biases,
$\Delta_{k,F}$ and $\delta_{k,F}$. Both the factorial cumulants
and the biases are related to the corresponding cumulants via
the linear relation Eqs. \ref{eq:cum_from_fac_cum} and \ref{eq:fac_cum_from_cum}.
The corrected factorial cumulants and the associated biases
are

\begin{align}
C_{2}^{corr} & =C_{2}[N]-\frac{\langle N\rangle^{2}}{\langle M\rangle^{2}}C_{2}[M]\\
C_{3}^{corr} & =C_{3}[N]-\frac{3C_{2}[M]C_{2}[N]\langle N\rangle}{\langle M\rangle^{2}}+\frac{3C_{2}[M]{}^{2}\langle N\rangle^{3}}{\langle M\rangle^{4}}-\frac{C_{3}[M]\langle N\rangle^{3}}{\langle M\rangle^{3}}\\
C_{4}^{corr} & = C_{4}[N]-\left(\frac{6C_{2}[N]\langle N\rangle^{2}\left(C_{3}[M]\langle M\rangle-3C_{2}[M]{}^{2}\right)}{\langle M\rangle^{4}}+\frac{4C_{2}[M]C_{3}[N]\langle N\rangle}{\langle M\rangle^{2}}\right.\\
 & \left.+\frac{3C_{2}[M]C_{2}[N]{}^{2}}{\langle M\rangle^{2}}+\frac{\langle N\rangle^{4}\left(-10C_{3}[M]C_{2}[M]\langle M\rangle+C_{4}[M]\langle M\rangle^{2}+15C_{2}[M]{}^{3}\right)}{\langle M\rangle^{6}}\right)
\end{align}

\begin{align}
\Delta_{2,F} & =\frac{\langle N\rangle^{2}}{\langle M\rangle^{2}}\bar{C}_{2}[M]\\
\Delta_{3,F} & =\bar{C}_{2}[M]\left(\frac{3C_{2}[N]\langle N\rangle}{\langle M\rangle^{2}}-\frac{3C_{2}[M]\langle N\rangle^{3}}{\langle M\rangle^{4}}\right)+\frac{\bar{C}_{3}[M]\langle N\rangle^{3}}{\langle M\rangle^{3}}\\
\Delta_{4,F} & =\bar{C}_{2}[M]\left(\frac{4C_{3}[N]\langle N\rangle}{\langle M\rangle^{2}}+\frac{3C_{2}[N]{}^{2}}{\langle M\rangle^{2}}-\frac{18C_{2}[M]C_{2}[N]\langle N\rangle^{2}}{\langle M\rangle^{4}}-\frac{4C_{3}[M]\langle N\rangle^{4}}{\langle M\rangle^{5}}+\frac{15C_{2}[M]{}^{2}\langle N\rangle^{4}}{\langle M\rangle^{6}}\right)\\
 & +\bar{C}_{3}[M]\left(\frac{6C_{2}[N]\langle N\rangle^{2}}{\langle M\rangle^{3}}-\frac{6C_{2}[M]\langle N\rangle^{4}}{\langle M\rangle^{5}}\right)+\frac{\bar{C}_{4}[M]\langle N\rangle^{4}}{\langle M\rangle^{4}}
\end{align}
For the scaled factorial cumulants we have 

\begin{align}
c_{2}^{corr} & =c_{2}[N]-\frac{\ave N}{\ave M} c_{2}[M]\\
c_{3}^{corr} & =c_{3}[N]-3\frac{\ave N}{\ave M}c_{2}[M]c_{2}[N]+\left(\frac{\ave N}{\ave M}\right)^{2}\left(3c_{2}[M]^{2}-c_{3}[M]\right)\\
c_{4}^{corr} & =c_{4}[N]-\frac{\ave N}{\ave M}\left(3c_{2}[M]c_{2}[N]{}^{2}+4c_{2}[M]c_{3}[N]\right)\\
 & +\left(\frac{\ave N}{\ave M}\right)^{2}\left(18c_{2}[M]^{2}c_{2}[N]-6c_{3}[M]c_{2}[N]\right)\\
 & +\left(\frac{\ave N}{\ave M}\right)^{3}\left(-15c_{2}[M]{}^{3}+10c_{2}[M]c_{3}[M]-c_{4}[M]\right)
\end{align}

\begin{align}
\delta_{2,F} & =\frac{\ave N}{\ave M}\bar{c}_{2}[M]\\
\delta_{3,F} & =3\frac{\ave N}{\ave M}c_{2}[N]\bar{c}_{2}[M]+\left(\frac{\ave N}{\ave M}\right)^{2}\left(\bar{c}_{3}[M]-3c_{2}[M]\bar{c}_{2}[M]\right)\\
\delta_{4,F} & =\frac{\ave N}{\ave M}\left(3c_{2}[N]{}^{2}\bar{c}_{2}[M]+4c_{3}[N]\bar{c}_{2}[M]\right)\\
 & +\left(\frac{\ave N}{\ave M}\right)^{2}\left(6c_{2}[N]\bar{c}_{3}[M]-18c_{2}[M]c_{2}[N]\bar{c}_{2}[M]\right)\\
 & +\left(\frac{\ave N}{\ave M}\right)^{3}\left(15c_{2}[M]{}^{2}\bar{c}_{2}[M]-6c_{2}[M]\bar{c}_{3}[M]-4c_{3}[M]\bar{c}_{2}[M]+\bar{c}_{4}[M]\right)
\end{align}

\section{Particle production through clusters}
\label{sec:clusters}
Let us assume that particles are produced via clusters and that each cluster further decays into two particles. Moreover,  clusters are generated from a Poisson distribution. As each cluster decays into two particles the probability of measuring $k$ particles is equivalent to measuring $k/2$ clusters and can be presented as:

\begin{equation}
    p\left(k;\langle N_{cl} \rangle\right) = e^{-\langle N_{cl}\rangle}\frac{\langle N_{cl}\rangle^{k/2}}{(k/2)!} 
\end{equation}

The corresponding moment generating function reads:

\begin{equation}
    M(t) = \sum_{k/2=0}^{\infty}e^{tk}e^{-\langle N_{cl}\rangle}\frac{\langle N_{cl}\rangle^{k/2}}{(k/2)!} = e^{\langle N_{cl}\rangle\left(e^{2t}-1\right)},
\end{equation}
where $\langle N_{cl}\rangle$ denotes mean number of clusters produced.

The cumulants of total particle number $k$ can be computed as:

\begin{equation}
    \kappa_{n}[k] = \left.\frac{\partial ln(M(t))}{dt}\right\rvert_{t = 0}
\end{equation}

For the first two cumulants one gets:

\begin{equation}
    \kappa_{1}[k] = 2\langle N_{cl}\rangle
    \label{cluster_kappa1}
\end{equation}

\begin{equation}
    \kappa_{2}[k] = 4\langle N_{cl}\rangle
    \label{cluster_kappa2}
\end{equation}

One clearly sees from Eqs.~\ref{cluster_kappa1} and~\ref{cluster_kappa2} that $\kappa_{1}(k) \ne \kappa_{2}(k)$, i.e the  total number of particles does not follow a Poisson distribution, although the clusters do. Moreover, one observes that particle production  through clusters enhances fluctuations.
In general, for clusters following a Poisson distribution and decaying into $m$ particles, the cumulants of total particle number can be written as:

\begin{equation}
    \kappa_{n}[k] = m^{n}\langle N_{cl}\rangle
    \label{cluster_kappam}
\end{equation}

\section{Mixed Cumulants}
\label{sec:mixed_cumulants}

Here we provide the relevant formulas for mixed cumulants. Given the
generating function, Eq. \ref{eq:cum_gen_wound}, the mixed cumulants
for particles of type $A$ and $B$ are given by (see Eq. \ref{eq:mixed_cum_wound})  
\begin{equation}
\kappa_{j,k}[A,B]=\left.\frac{\partial^{j}\partial^{k}}{\partial t_{A}\partial t_{B}}g\left(t_{A},t_{B}\right)\right|_{t_{A}=t_{B}=0}
\end{equation}
The explicit formulas for the four lowest-order mixed cumulants are: 
\begin{align}
\kappa_{1,1}[A,B] & =\bar{\kappa}_{1,1}[A,B]+\langle A\rangle\langle B\rangle\frac{\kappa_{2}\left[N_{W}\right]}{\left\langle N_{W}\right\rangle {}^{2}}\\
\kappa_{2,1}[A,B] & =\bar{\kappa}_{2,1}[A,B]+\left(2\langle A\rangle\bar{\kappa}_{1,1}[A,B]+\langle B\rangle\bar{\kappa}_{2,0}[A,B]\right)\frac{\kappa_{2}\left[N_{W}\right]}{\left\langle N_{W}\right\rangle {}^{2}}+\langle A\rangle^{2}\langle B\rangle\frac{\kappa_{3}\left[N_{W}\right]}{\left\langle N_{W}\right\rangle {}^{3}}\\
\kappa_{1,2}[A,B] & =\bar{\kappa}_{1,2}[A,B]+\left(\langle A\rangle\bar{\kappa}_{0,2}[A,B]+2\langle B\rangle\bar{\kappa}_{1,1}[A,B]\right)\frac{\kappa_{2}\left[N_{W}\right]}{\left\langle N_{W}\right\rangle {}^{2}}+\langle A\rangle\langle B\rangle^{2}\frac{\kappa_{3}\left[N_{W}\right]}{\left\langle N_{W}\right\rangle {}^{3}}\\
\kappa_{2,2}[A,B] & =\bar{\kappa}_{2,2}[A,B]+\left(\langle
                    A\rangle^{2}\bar{\kappa}_{0,2}[A,B]+4\langle A\rangle\langle
                    B\rangle\bar{\kappa}_{1,1}[A,B]+\langle
                    B\rangle^{2}\bar{\kappa}_{2,0}[A,B]\right)\frac{\kappa_{3}\left[N_{W}\right]}{\left\langle
                    N_{W}\right\rangle {}^{3}} \nonumber\\
 & +\left(2\langle A\rangle\bar{\kappa}_{1,2}[A,B]+2\langle
   B\rangle\bar{\kappa}_{2,1}[A,B]+2\bar{\kappa}_{1,1}[A,B]{}^{2}+\bar{\kappa}_{0,2}[A,B]\bar{\kappa}_{2,0}[A,B]\right)\frac{\kappa_{2}\left[N_{W}\right]}{\left\langle
   N_{W}\right\rangle {}^{2}} \nonumber\\
 & +\langle A\rangle^{2}\langle B\rangle^{2}\frac{\kappa_{4}\left[N_{W}\right]}{\left\langle N_{W}\right\rangle {}^{4}}
\end{align}
where, analogous to the notation for the regular cumulants, $\Kb{j,k}[A,B]$
denotes the mixed cumulant for constant number of wounded nucleons
$\ave{N_{W}}$. Note, that $\kappa_{j,0}[A,B]=\kappa_{j}[A]$ and
$\kappa_{0,j}[A,B]=\kappa_{j}[B]$ correspond to the regular cumulant
for particles of type $A$ and $B$ respectively. The firs order mixed
cumulant, $\kappa_{1,1}[A,B]=cov[A,B]$ is also referred to as the
co-variance between the distributions of particles $A$ and $B$.
In order to obtain the corrected mixed cumulants we proceed in the
same fashion as for the regular cumulant. We express the terms involving
cumulants of the wounded nucleons, $\kappa_{i}\left[N_{W}\right]/\ave{N_{W}}^{i}$
in terms of the factorial cumulants of the multiplicity distribution
(See Eqs. \ref{eq:k_wound_2}-\ref{eq:k_wound_4}) and solve for the
the mixed cumulants with fixed number of wounded nucleons, $\Kb{j,k}[A,B]$.
Again, the results are given in the form
\begin{align}
\Kb{j,k}[A,B]=\kappa_{j,k}^{corr}[A,B]+\Delta_{j,k}
\end{align}
where $\kappa_{j,k}^{corr}[A,B]$ are the cumulants including the
measurable corrections and $\Delta_{j,k}$ are the corresponding biases
due to quantities which are not directly measurable. 
\begin{align}
\kappa_{1,1}^{corr}[A,B] & =\kappa_{1,1}[A,B]-\frac{\langle A\rangle\langle B\rangle}{\langle M\rangle^{2}}C_{2}[M]\\
\Delta_{1,1} & =\frac{\langle A\rangle\langle B\rangle}{\langle M\rangle^{2}}\bar{C}_{2}[M]
\label{eq:kappa_11}
\\
\kappa_{2,1}^{corr}[A,B] & =\kappa_{2,1}[A,B]-\frac{\langle B\rangle C_{2}[M]\bar{\kappa}_{2,0}[A,B]}{\langle M\rangle^{2}}-\frac{2\langle A\rangle C_{2}[M]\kappa_{1,1}[A,B]}{\langle M\rangle^{2}}+\frac{\langle A\rangle^{2}\langle B\rangle\left(2C_{2}[M]{}^{2}-\langle M\rangle C_{3}[M]\right)}{\langle M\rangle^{4}}\\
\Delta_{2,1} & =\frac{1}{\langle M\rangle^{4}}\left[2\langle A\rangle\langle M\rangle^{2}\bar{C}_{2}[M]\kappa_{1,1}[A,B]+\langle B\rangle\langle M\rangle^{2}\bar{C}_{2}[M]\bar{\kappa}_{2,0}[A,B]\right.\nonumber \\
 & \left.-\langle A\rangle^{2}\langle B\rangle\left(\bar{C}_{2}[M]\left(\bar{C}_{2}[M]+C_{2}[M]\right)-\langle M\rangle\bar{C}_{3}[M]\right)\right]\\
\kappa_{1,2}^{corr}[A,B] & =\kappa_{1,2}[A,B]-\frac{\langle A\rangle C_{2}[M]\bar{\kappa}_{0,2}[A,B]}{\langle M\rangle^{2}}-\frac{2\langle B\rangle C_{2}[M]\kappa_{1,1}[A,B]}{\langle M\rangle^{2}}+\frac{\langle B\rangle^{2}\langle A\rangle\left(2C_{2}[M]{}^{2}-\langle M\rangle C_{3}[M]\right)}{\langle M\rangle^{4}}\\
\Delta_{1,2} & =\frac{1}{\langle M\rangle^{4}}\left[2\langle B\rangle\langle M\rangle^{2}\bar{C}_{2}[M]\kappa_{1,1}[A,B]+\langle A\rangle\langle M\rangle^{2}\bar{C}_{2}[M]\bar{\kappa}_{0,2}[A,B]\right.\nonumber \\
 & \left.-\langle A\rangle\langle B\rangle^{2}\left(\bar{C}_{2}[M]\left(\bar{C}_{2}[M]+C_{2}[M]\right)-\langle M\rangle\bar{C}_{3}[M]\right)\right]\\
\kappa_{2,2}^{corr}[A,B] & =\kappa_{2,2}[A,B]+\frac{2\langle A\rangle^{2}C_{2}[M]{}^{2}\bar{\kappa}_{0,2}[A,B]}{\langle M\rangle^{4}}+\frac{2\langle B\rangle^{2}C_{2}[M]{}^{2}\bar{\kappa}_{2,0}[A,B]}{\langle M\rangle^{4}}\nonumber \\
 & -\frac{C_{2}[M]\bar{\kappa}_{0,2}[A,B]\bar{\kappa}_{2,0}[A,B]}{\langle M\rangle^{2}}-\frac{\langle A\rangle^{2}C_{3}[M]\bar{\kappa}_{0,2}[A,B]}{\langle M\rangle^{3}}-\frac{\langle B\rangle^{2}C_{3}[M]\bar{\kappa}_{2,0}[A,B]}{\langle M\rangle^{3}}\nonumber \\
 & +\frac{12\langle A\rangle\langle B\rangle C_{2}[M]{}^{2}\kappa_{1,1}[A,B]}{\langle M\rangle^{4}}-\frac{2C_{2}[M]\kappa_{1,1}[A,B]{}^{2}}{\langle M\rangle^{2}}-\frac{2\langle A\rangle C_{2}[M]\kappa_{1,2}[A,B]}{\langle M\rangle^{2}}\nonumber \\
 & -\frac{2\langle B\rangle C_{2}[M]\kappa_{2,1}[A,B]}{\langle M\rangle^{2}}-\frac{4\langle A\rangle\langle B\rangle C_{3}[M]\kappa_{1,1}[A,B]}{\langle M\rangle^{3}}-\frac{10\langle A\rangle^{2}\langle B\rangle^{2}C_{2}[M]{}^{3}}{\langle M\rangle^{6}}\nonumber \\
 & +\frac{8\langle A\rangle^{2}\langle B\rangle^{2}C_{3}[M]C_{2}[M]}{\langle M\rangle^{5}}-\frac{\langle A\rangle^{2}\langle B\rangle^{2}C_{4}[M]}{\langle M\rangle^{4}}\\
\Delta_{2,2} & =\bar{C}_{2}[M]{}^{2}\left(-\frac{\langle A\rangle^{2}\bar{\kappa}_{0,2}[A,B]}{\langle M\rangle^{4}}-\frac{\langle B\rangle^{2}\bar{\kappa}_{2,0}[A,B]}{\langle M\rangle^{4}}+\frac{3C_{2}[M]\langle A\rangle^{2}\langle B\rangle^{2}}{\langle M\rangle^{6}}\right)\nonumber \\
 & +\bar{C}_{2}[M]\left(-\frac{C_{2}[M]\langle A\rangle^{2}\bar{\kappa}_{0,2}[A,B]}{\langle M\rangle^{4}}-\frac{C_{2}[M]\langle B\rangle^{2}\bar{\kappa}_{2,0}[A,B]}{\langle M\rangle^{4}}+\frac{\bar{\kappa}_{0,2}[A,B]\bar{\kappa}_{2,0}[A,B]}{\langle M\rangle^{2}}\right.\nonumber \\
 & \;\;-\frac{2\bar{C}_{3}[M]\langle A\rangle^{2}\langle B\rangle^{2}}{\langle M\rangle^{5}}-\frac{12C_{2}[M]\langle A\rangle\langle B\rangle\kappa_{1,1}[A,B]}{\langle M\rangle^{4}}+\frac{2\langle A\rangle\kappa_{1,2}[A,B]}{\langle M\rangle^{2}}+\frac{2\kappa_{1,1}[A,B]{}^{2}}{\langle M\rangle^{2}}\nonumber \\
 & \left.\;+\frac{2\langle B\rangle\kappa_{2,1}[A,B]}{\langle M\rangle^{2}}+\frac{6C_{2}[M]{}^{2}\langle A\rangle^{2}\langle B\rangle^{2}}{\langle M\rangle^{6}}-\frac{2C_{3}[M]\langle A\rangle^{2}\langle B\rangle^{2}}{\langle M\rangle^{5}}\right)\nonumber \\
 & +\bar{C}_{3}[M]\left(\frac{\langle A\rangle^{2}\bar{\kappa}_{0,2}[A,B]}{\langle M\rangle^{3}}+\frac{\langle B\rangle^{2}\bar{\kappa}_{2,0}[A,B]}{\langle M\rangle^{3}}+\frac{4\langle A\rangle\langle B\rangle\kappa_{1,1}[A,B]}{\langle M\rangle^{3}}-\frac{4C_{2}[M]\langle A\rangle^{2}\langle B\rangle^{2}}{\langle M\rangle^{5}}\right)\nonumber \\
 & +\frac{\bar{C}_{2}[M]{}^{3}\langle A\rangle^{2}\langle B\rangle^{2}}{\langle M\rangle^{6}}+\frac{\bar{C}_{4}[M]\langle A\rangle^{2}\langle B\rangle^{2}}{\langle M\rangle^{4}}
\end{align}

\bibliographystyle{apsrev4-1}
\bibliography{paper}{}
\end{document}

%% file: macros.tex
\ifx\ave\undefined 
\global\long\def\ave#1{\left\langle #1 \right\rangle }%
 \fi

\ifx\absol\undefined 
\global\long\def\absol#1{\left| #1 \right|}%
 \fi

\ifx\mev\undefined 
\global\long\def\mev{{\rm \, MeV}}%
 \fi

\ifx\gev\undefined 
\global\long\def\gev{{\rm \, GeV}}%
 \fi

\ifx\tev\undefined 
\global\long\def\tev{{\rm \, TeV}}%
 \fi

\ifx\dpp\undefined 
\global\long\def\dpp#1{\frac{d^{3}#1}{(2\pi)^{3}}}%
 \fi

\ifx\mh\undefined 
\global\long\def\mh{\hat{\mu}}%
 \fi